\documentclass[twocolumn,10pt]{asme2ej}

\usepackage{epsfig}
\usepackage{amsmath}
\usepackage{hyperref}
\usepackage{graphicx}
\usepackage{subfigure}
\usepackage{verbatim}
\usepackage{ulem}
\usepackage{enumerate}

\graphicspath{{images/}}

\usepackage{color}
\usepackage{ulem}
\newcommand{\rev}[2]{#2}

\title{Memory-Efficient Modeling and Slicing of Large-Scale Adaptive Lattice Structures}

\author{\hspace{-10pt}
Shengjun Liu$^1$, Tao Liu$^1$, Qiang Zou$^{2,5}$, Weiming Wang$^{3,4}$, Eugeni L. Doubrovski$^4$, Charlie C.L. Wang$^5$\thanks{Corresponding Author; Email: changling.wang@manchester.ac.uk} 
\\
\affiliation{
$^1$School of Mathematics and Statistics, Central South University, China\\
$^2$State Key Laboratory of CAD\&CG, Zhejiang University, Hangzhou, China\\
$^3$School of Mathematical Sciences, Dalian University of Technology, China\\
$^4$Faculty of Industrial Design Engineering, Delft University of Technology, The Netherlands\\
$^5$Department of Mechanical, Aerospace and Civil Engineering, University of Manchester, UK
}
}

\begin{document}

\maketitle

\begin{abstract}
{\it
Lattice structures have been widely used in various applications of additive manufacturing due to its superior physical properties. If modeled by triangular meshes, a lattice structure with huge number of struts would consume massive memory. This hinders the use of lattice structures in large-scale applications (e.g., to design the interior structure of a solid with spatially graded material properties). To solve this issue, we propose a memory-efficient method for the modeling and slicing of adaptive lattice structures. A lattice structure is represented by a weighted graph where the edge weights store the struts' radii. When slicing the structure, its solid model is locally evaluated through convolution surfaces and in a streaming manner. As such, only limited memory is needed to generate the toolpaths of fabrication. Also, the use of convolution surfaces leads to natural blending at intersections of struts, which can avoid the stress concentration at these regions. We also present a computational framework for optimizing supporting structures and adapting lattice structures with prescribed density distributions. The presented methods have been validated by a series of case studies with large number (up to 100M) of struts to demonstrate its applicability to large-scale lattice structures.
}
\end{abstract}



\section{Introduction}
Additive manufacturing has enabled the fabrication of objects with highly complicated shapes and structures. Recently, increasing attention has been drawn towards modeling interior structures of 3D models rather than the exterior appearance \cite{martinez2016procedural,martinez2017orthotropic,kuipers2019CrossFill}. One important type of interior structures is the lattice structure, which consists of interconnected struts. Such structures are lightweight, yet have superior mechanical properties \cite{qin2017mechanics}. When such lightweight structures are used in vehicles, less energy will be consumed. Moreover, with carefully designed density distributions, spatially graded material properties can be realized at different regions of a 3D printed lattice structure, which therefore introduces varied mechanical properties in the part. 

When using lattice structures to realize spatially graded material properties, the number of struts can be huge. In such a scenario, if triangular meshes (i.e., the de facto standard representation format in 3D printing) are to be used to represent lattice structures, the memory consumption can be extremely high. This poses a significant computational challenge in applications of large-scale, adaptive lattices structures. The previously developed out-of-core modeling algorithms (e.g.~\cite{rosen2006lattice,chen2008LDNI}) for lattice structures could handle uniform or periodical lattice structures but becomes difficult for processing large-scale adaptive lattice structures. There is also prior research  \cite{chougrani2017lattice} seeking to reduce the number of facets in triangulating a lattice structure. However, when a tremendous number of struts are involved to model lattice structures, the method will generate too many triangles to run on a general computer.

To address the aforementioned challenge of large-scale lattice structures in AM, we propose a memory-efficient method for modeling and slicing adaptive lattice structures in large-scales. A lattice structure is represented by a weighted graph with edges representing struts, nodes representing intersections of the struts, and edge weights specifying radii of the corresponding struts. The corresponding solid of the lattice structure is locally defined using convolution surfaces with compactly supported kernel functions. In this way, the solid's  boundary surface will only be locally generated when needed. This yields a highly scalable representation for adaptive lattice structures. In addition, solid models based on convolution surfaces have naturally blended shape at the intersections of struts, which can avoid the stress concentration at these regions. Based on the above representation scheme, a streaming based slicing algorithm for 3D printing is then developed to demonstrate the scalability of our method -- models with a massive number (up to 100M) of struts can be successfully handled. 

Based on the scalable modeling method, an algorithm for constructing a lattice structure model in accordance with a prescribed density distribution is also presented in this paper. The lattice structure is initialized with edges and nodes of a tetrahedral mesh generated by the method presented in Ref.~\cite{si2015tetgen}. Then, the density in the initial lattice structure is adjusted to match with the prescribed density in two steps as tetrahedra subdivision and strut radius adjustment. To facilitate the printing process, the lattice structure is further optimized to be self-supported for additive manufacturing.

The technical contributions of our work are:
\begin{enumerate}
\item A new memory-efficient representation of lattice structures that can be employed to solve large-scale modeling problem of spatially graded material properties.

\item A slicing algorithm in streaming mode to realize the fabrication of lattice structures with a large number of struts.

\item \rev{We demonstrate the functionality of our method on developing a computational framework for inverse design -- constructing a lattice structure that matches a prescribed distribution of density.}{A computational framework to reversely design a lattice structure that matches a prescribed density distribution and achieves the optimal self-supporting. This framework, when combined with the above two contributions, provides a comprehensive and practical pipeline for modeling and slicing lattice structures.}
\end{enumerate}

The rest of this paper is organized as follows. After reviewing the related works in Section \ref{secReview}, we present the modeling method of large-scale lattice structures in Section \ref{secModeling}, which is followed by introducing a streaming-based slicing algorithm in Section \ref{secSlicing}. The computational framework for constructing a lattice structure according to the required density distribution is then presented in Section \ref{secDensity}. The effectiveness of our approach is evaluated in Section \ref{secResult} and our paper ends with the conclusion.

\section{Related Works}\label{secReview}
In literature, there are numerous techniques of modeling complicated geometry for additive manufacturing applications. Here we only study the most relevant works. More comprehensive surveys can be found in \cite{gao2015status,livesu20173d,Leung2019}.

\rev{Different from the techniques that generating tool-paths for densely filling the interior of a contour on each layer (e.g., \cite{ding2014tool,zhao2016connected,steuben2016implicit}), more and more researches are conducted to explore optimal infill-structures of 3D printed models to generate tailor-made physical properties.}{Different from conventional methods (e.g., \cite{ding2014tool,zhao2016connected,steuben2016implicit}) that generate a full infill inside a part, more and more research studies are being conducted to design distributed infill structures for functionally tailored 3D printing.} Examples include fractal-like space filling curves / surfaces \cite{kuipers2019CrossFill,kumar2009fractal}, adaptive rhombic grid \cite{wu2016self,lee2017block}, Voronoi diagram based structure \cite{martinez2016procedural,lu2014build,lee2018support,Stankovic2020}, beam-like structure \cite{wang2013cost,zhang2015medial}, periodic tiles \cite{schumacher2015microstructures,panetta2015elastic}, procedural periodic tiles \cite{fryazinov2013multi,martinez2017orthotropic}, and voxel-based structure \cite{yang2018computing}. Among these approaches, only the approaches presented in \cite{martinez2016procedural} and \cite{kuipers2019CrossFill} provide a method to construct an infill-structure matching the given density field. However, the demand of self-supporting is not considered in \cite{martinez2016procedural}, and thin-shell structures are employed in \cite{kuipers2019CrossFill} which is not able to achieve high sparsity as the lattice structures studied in this work. 

In general, the inner surfaces of a model are represented by two-manifold triangular meshes, which can be obtained from different representations, 
such as voxels in \cite{yang2018computing}, tetrahedral meshes in \cite{christiansen2015automatic}, rhombic cells in \cite{wu2016self,lee2017block}, elliptic cylunders in \cite{lee2018support}
and extended distance-fields in \cite{martinez2016procedural,martinez2017orthotropic}. In this paper, we introduce a skeletonal representation of lattice structures by only storing the set of nodes and the edge-connectivity of a graph as skeletons. The solid can be efficiently and effectively evaluated from these skeletons by convolution surface with compactly supported kernel functions. Different from distance-fields, the formulation of convolution surface provides highly smooth surfaces at the intersected regions of struts. This can avoid the stress concentration at those regions with sharp creases. In addition, due to local kernel functions used in the convolution surface, the computation in slicing algorithm can be evaluated in a streaming manner -- i.e., with high scalability.

When additive manufacturing is applied to fabricate models with complicated geometry, supporting structures need to be added below the part with large overhang \rev{}{\cite{stava2012stress,ou2016cilllia}}. In general, the additional support will lead to the problems of hard-to-remove, surface damage and additional cost of material and fabrication time. For the 3D printed models of spatially graded density, the additional support may change the designed density distribution. In literature, many approaches have been developed to reduce the usage of supporting structures. For example, some methods tried to compute an optimal printing direction to reduce the influence of supporting structure \cite{vanek2014clever,zhang2015perceptual}. Different supporting structures are designed to reduce the volume of material usage, such as the tree-structures proposed in \cite{vanek2014clever} and the bridge structures proposed in \cite{dumas2014bridging}. Moreover, approaches have also been developed to generate completely self-supported infill structures (ref.~\cite{wu2016self,lee2017block,lee2018support,yang2018computing,kuipers2019CrossFill}). As a design tool, we propose to conduct the strategy employed in \cite{hu2015support} to deform a model to reduce the demand of support.


\begin{figure*}[t]
\centering
\includegraphics[width=\linewidth]{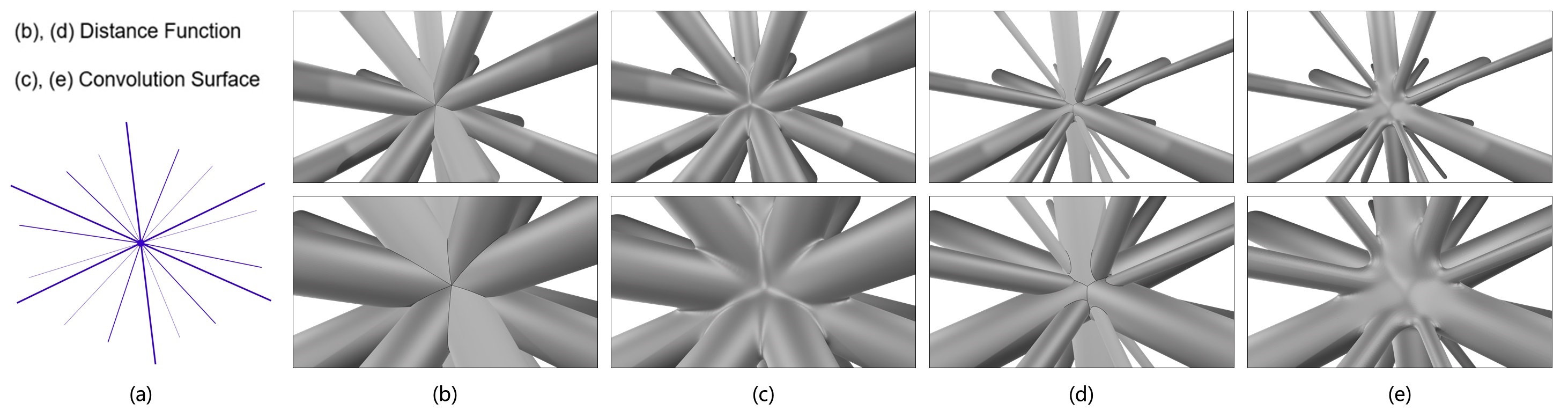}
\caption{For a given skeleton model as shown in (a), the solid models generated by distance field have sharp creases -- see (b) by using uniform radius and (d) by using different radii for different edges of the skeleton. Differently, solids with smooth surfaces can be constructed by our approach in (c) and (e).}
\label{fig:Crease}
\end{figure*}

\section{Modeling of Lattice Structure}\label{secModeling}
In this section, we will first present the representation of lattice structures and then formulate its implicit solid by convolution surface with compactly supported kernel functions. After that, we study shape error at the joint regions which may have over-blending problems. 

\subsection{Implicit Solid Representation}\label{subsecImplicitSolid}
A lattice structure is represented as a graph of interconnected skeletons $\Omega$, which is stored as a complex-based data structure $\Omega=(\mathcal{V}, \mathcal{E})$ with a set of nodes and a set of edges being a simplified version of the data structure for general non-manifold objects \cite{wang2003nonmanifold}. For each  $\mathbf{v}_i \in \mathcal{V}$, it defines the position of a node as $\mathbf{v}_i \in \Re^3$. For an edge $e_j \in \mathcal{E}$, it is represented as a pair of vertices associated with the radius of the edge's corresponding strut as $e_j=(\mathbf{v}_s, \mathbf{v}_e, r_j)$. The solid of a lattice structure is formulated as an implicit surface defined around the skeletons. 

Given the skeleton representation $\Omega$ of a lattice structure, its corresponding solid is formulated as
\begin{equation}
\mathcal{S}(\Omega)=\{ \mathbf{p} \; | \; F(\mathbf{p}) \leq 0 \; (\forall \mathbf{p} \in \Re^3) \},
\end{equation}
where $F(\cdot)$ is an implicit function returning the value proportional to the distance between $\mathbf{p}$ and $\Omega$. A straightforward solution is to use an offset \textit{distance function} as
\begin{equation}
F(\mathbf{p}) = - D + \min_{\forall \mathbf{q} \in \Omega} \left\| \mathbf{q} - \mathbf{p}\right\|   
\end{equation}
with $D$ \rev{be}{being} assigned as $D = r_j$ when the closest point of $\mathbf{p}$ is located on the edge $e_j$. This definition based on distance function has two problems:
\begin{enumerate}
\item The closest point of a query $\mathbf{p}$ is not unique -- there may \rev{have}{be} multiple closest points. As a consequence, the value of $D$ (i.e., the function value of $F(\mathbf{p})$) is not well-defined when different radii $r_j$s are assigned to different edges in $\Omega$.

\item Sharp creases are formed at the boundary surface of solid due to the discontinuity of the distance function (see Fig.\ref{fig:Crease}(b) and (d)). Concentrated stresses can be easily generated in these regions. 
\end{enumerate}
A formulation of $F(\cdot)$ based on convolution surface is employed in our framework, which can essentially solve these two problems. 
\begin{equation}
    F(\mathbf{p}) = -C + \int_{V}{h(\mathbf{x})f(\mathbf{p}-\mathbf{x})dV} =-C +(f \otimes h)(\mathbf{p}), 
\end{equation}
where $h:\Re^3 \mapsto \Re$ is a geometric function representing the skeleton $\Omega$
\begin{align}
\centering
h(\mathbf{x})=\left\{ \begin{array}{ll}
1, & \mathbf{x}\in \Omega \\
0, & \mathrm{otherwise}
\end{array}\right.,
\end{align}
and $C$ is a constant isovalue defined according to the radii defined on the skeleton edges. 

\begin{figure}[!t]
\centering
\subfigure[Along an edge]{\label{fig:FieldCompA}
\includegraphics[width=0.25\linewidth]{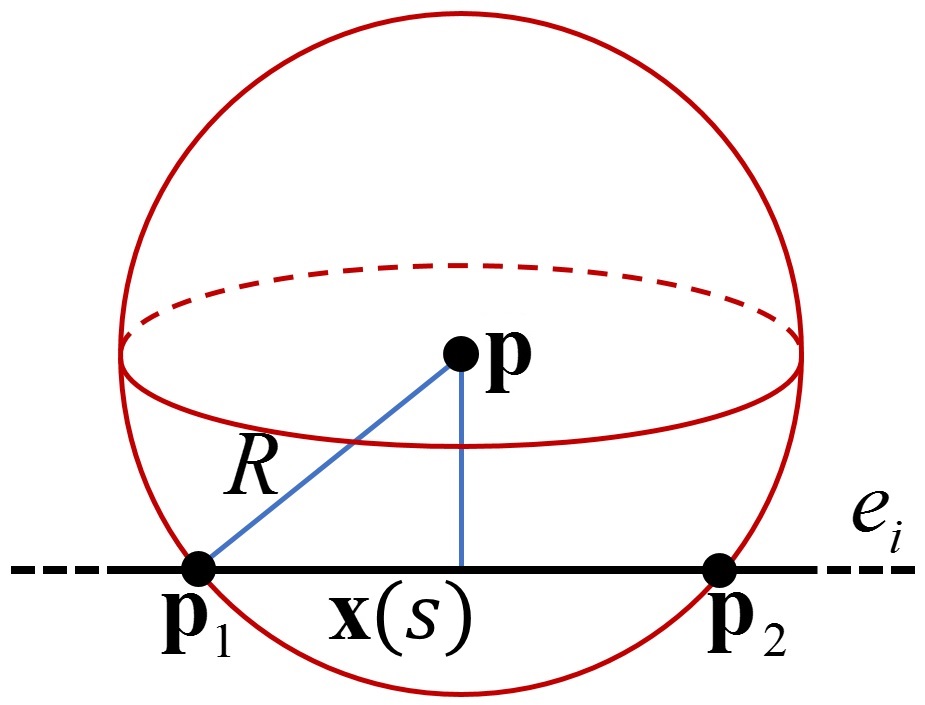}
}
\subfigure[Adaptive strut radius by changing $r_i$]{\label{fig:FieldCompB}
\includegraphics[width=0.65\linewidth]{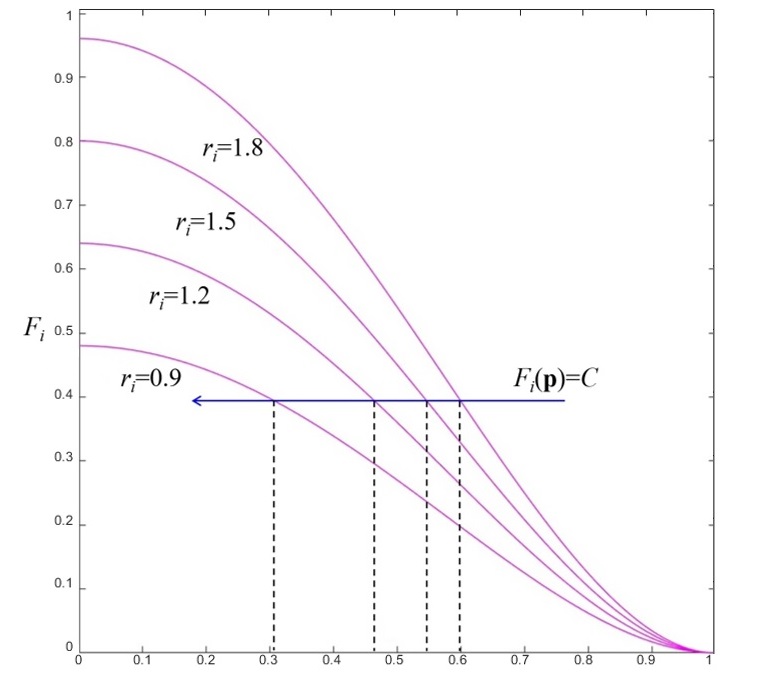}
}
\caption{Illustrations for computing the field-value of $F_i(\mathbf{p})$.}
\label{fig:FieldComp}
\end{figure}
\subsection{Representation with Compactly Supported Kernels}
To enable the modeling of lattice structures in large-scale, compactly supported kernels are employed in our method. \rev{}{The local support leads to local modification property, which is important for the fast computation of  slicing in Section \ref{secSlicing} and also the later density control in Section \ref{secDensity}.} Specifically, \rev{}{a quartic polynomial kernel function}
\begin{align}
\label{eqFuncImplicitKernel}
\centering
f(\mathbf{p}-\mathbf{x})=\left\{ \begin{array}{ll}
\left(1-{\| \mathbf{p}-\mathbf{x} \|^2}/{R^2} \right)^2, & \|\mathbf{p}-\mathbf{x}\| < R \\
0, & \mathrm{otherwise}
\end{array}\right.
\end{align}
with $R$ being the support-size \rev{of a kernel function}{is adopted and its evaluation is very simple and efficient, as already noted in \cite{sherstyuk1999kernel}}. Moreover, a local weight is assigned to each edge to specify the radius of its corresponding strut\rev{}{, as discussed in \cite{hubert2012convolution}}. Therefore, the convolution surface can be defined as
\begin{equation}\label{eqFuncImplicit}
F(\mathbf{p})= -C + \sum_{e_i \in \Omega} r_i \int_{\mathbf{x} \in e_i} f(\mathbf{p}-\mathbf{x}) d e_i
\end{equation}
which can \rev{}{be} efficiently evaluated. Specifically, for a query point $\mathbf{p}$, nonzero terms in Eq.(\ref{eqFuncImplicit}) only contains the edges with distance to $\mathbf{p}$ less than $R$. Moreover, the convolution integral along an edge can be computed by an analytical close-from. Details are given below.

Without loss of generality, it is assumed that the line of $e_i$ intersects with the sphere (centered at $\mathbf{p}$ and radius $R$) at two points $\mathbf{p}_1$ and $\mathbf{p}_2$ (see Fig.\ref{fig:FieldCompA}). The valid segment of $e_i$ inside the sphere can be defined in a parametric form as
\begin{center}
$\mathbf{x}(s) = (1-s)\mathbf{p}_1 + s \mathbf{p}_2 \qquad (\forall s \in [s_1,s_2])$  
\end{center}
with 
\begin{center}
$s_1=\max\{0,(\mathbf{v}_s-\mathbf{p}_1)\cdot (\mathbf{p}_2-\mathbf{p}_1) / \|\mathbf{p}_2-\mathbf{p}_1\|^2 \}$,
$s_2=\min\{1,(\mathbf{v}_e-\mathbf{p}_1)\cdot (\mathbf{p}_2-\mathbf{p}_1) / \|\mathbf{p}_2-\mathbf{p}_1\|^2 \}$.
\end{center}
As a result, the field value at point $\mathbf{p}$ contributed by $e_i$ can be computed as
\begin{align}
\begin{array}{ll}
\rev{F_i(\mathbf{p})}{F_{e_i}(\mathbf{p})} & = r_i\int_{e_i}f(\mathbf{p}-\mathbf{x}) d{e_i} \\
 & = r_i\int_{s_1}^{s_2} \left(1-{\| \mathbf{p}-\mathbf{x}(s) \|^2}/{R^2} \right)^2 ds \\
 & = \frac{r_i}{15 R^{4}} \left(3 l^{4} s^5- 15 a l^{2} s^{4}+ 20a^2 s^3 \right) \Big|_{s_1}^{s_2}
\end{array}
\end{align}
with $l=\|\mathbf{p}_{2}-\mathbf{p}_{1}\|$ and $a=(\mathbf{p} - \mathbf {p}_{1}) \cdot (\mathbf{p}_{2}- \mathbf{p}_{1})$.

When the isovalue $C$ in Eq.(\ref{eqFuncImplicit}) is fixed, we can adjust the value of $r_i$ to change the radius of strut generated from each skeleton edge $e_i$. While decreasing the value of $r_i$, a point with the same isovalue will be closer to $e_i$ thus reducing the strut radius. As illustrated in Fig.\ref{fig:FieldCompB}, we can reduce the radius of strut along the direction of the blue arrow by using smaller value of $r_i$ for each kernel function \rev{$F_i(\mathbf{p})$}{$F_{e_i}(\mathbf{p})$}. An example is given in Fig.\ref{fig:Crease}(e) that represents a solid having the same skeleton as the one shown in Fig.\ref{fig:Crease}(c) but using different radii for struts. 

In our formulation, the support size of each kernel function is not changed even when using different values of $r_i$s for different struts. The support size $R$ for all kernel functions serves as the upper bound for the radii that can be realized by convolution. As a consequence, users can select the value of $R$ by the range of strut radii that they wish to obtain (e.g., $1.5 \times$ the maximal radius used in our implementation).

\begin{figure}[t]
\centering
\includegraphics[width=\linewidth]{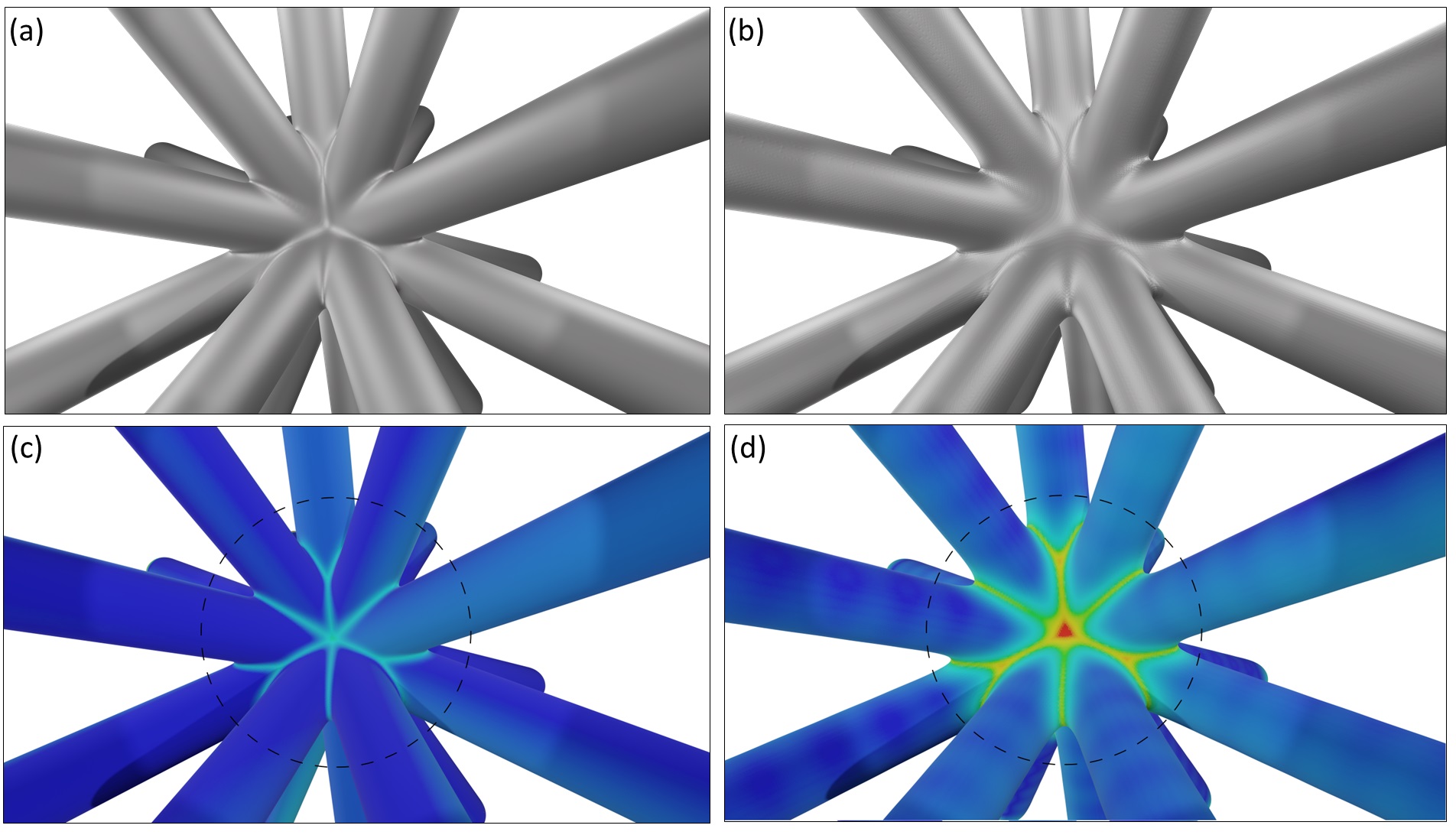}
\caption{Comparison of solids generated by our approach (a) and the Gaussian kernel (b) - both for the skeletons shown in Fig.\ref{fig:Crease}(a). It can clearly be observed that more over-blending artifacts are generated on the result of Gaussian kernel. The color maps in (c) and (d) are used to visualize the distance between implicit surfaces and the `ideal' solid as the union of cylinders.
}
\label{fig:figGaussian}
\end{figure}

\begin{figure}[t]
\centering
\includegraphics[width=\linewidth]{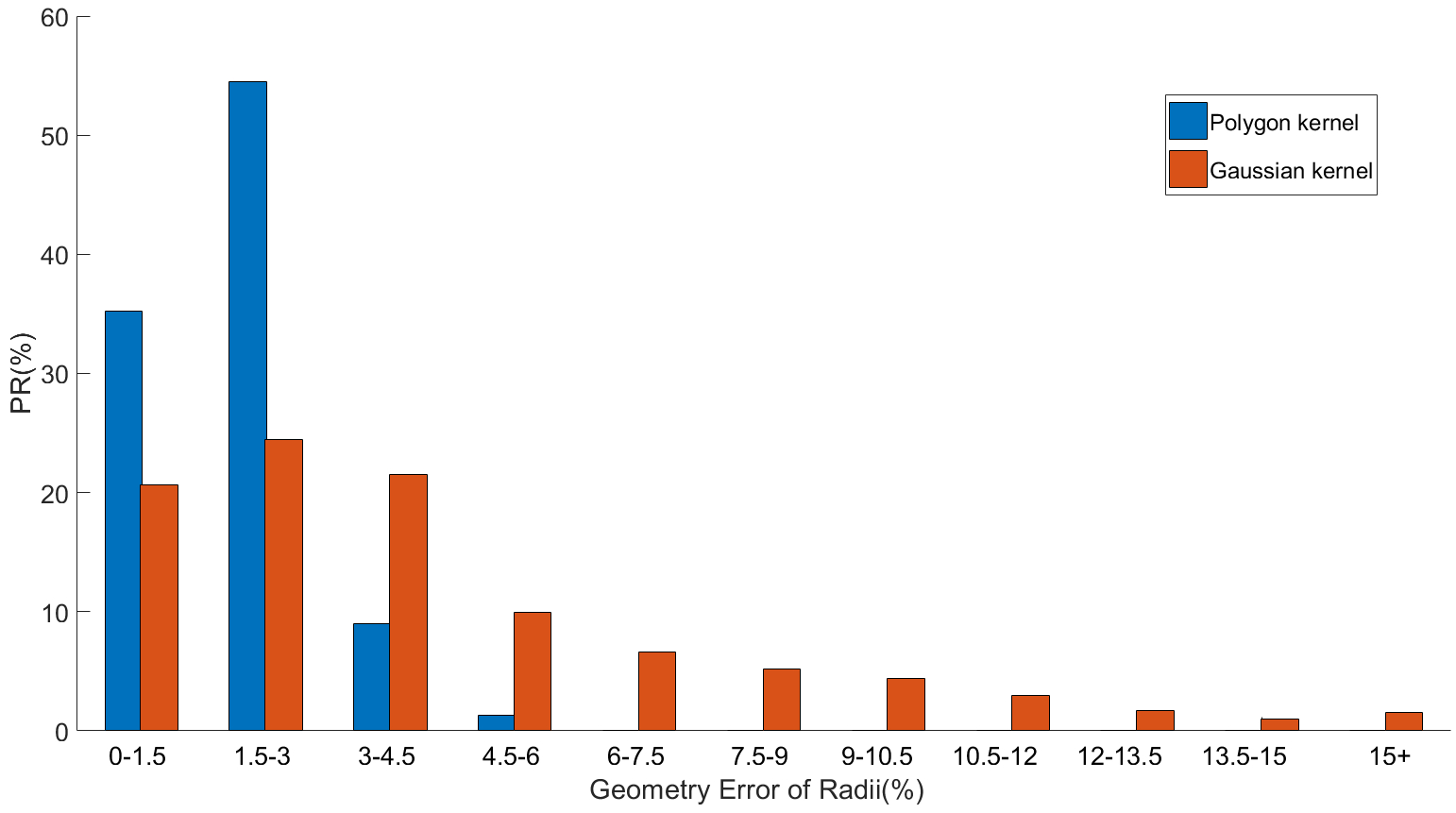}
\caption{
To visualize the surface distance errors, we generate sample points in a spherical region with radius as $2.83R$ -- the circled region shown in Fig.\ref{fig:figGaussian}(c) and (d). Histogram of distances between sample points to the `ideal' solid is given. 
For the result with Gaussian kernel, around $40\%$ sample points have the distance more than $6\% r$. Differently, our result has no sample point with distance more than $6\% r$. 
}
\label{fig:GaussianLocalComparebar}
\end{figure}

Comparing to the convolution solid generated by globally defined kernels such as Gaussian \cite{Tang2019,jin2002convolution} or by \textit{Fast Fourier Transform} (FFT) \cite{Nelaturi2015}, the benefit of convolution by compactly supported kernel functions is twofold.
\begin{enumerate}
\item The evaluation of implicit solid can be conducted in a out-of-core manner. For evaluating the function value at a point $\mathbf{p} \in \Re^3$, only those kernel functions with the distance between $\mathbf{p}$ and their centers less than $R$ will involve. Therefore, large-scale lattice structures can be efficiently modeled and sliced in our approach (see the streaming-mode slicing algorithm presented in the Section \ref{secSlicing}).

\item A convolution surface constructed by compactly supported kernels will generate less over-blending artifacts in the region with many overlapped kernels, \rev{(e.g., the joint shown in Fig.\ref{fig:Crease} associated with 18 skeletons). Detail comparison with Gaussian based convolution has been given in Figs.\ref{fig:figGaussian} and \ref{fig:GaussianLocalComparebar}}{as demonstrated by the comparison in Figs.\ref{fig:figGaussian} and \ref{fig:GaussianLocalComparebar}. The reason for this advantage is that over-blending is often caused by including unwanted contribution from nearby kernels when computing convolutions, and the limited range of compactly supported kernels can prevent such unwanted contribution to some degree.}  
\end{enumerate}
Both advantages are very important for efficiently and effectively modeling large-scale adaptive lattice structures.


\begin{figure*}
\centering
\includegraphics[width=\linewidth]{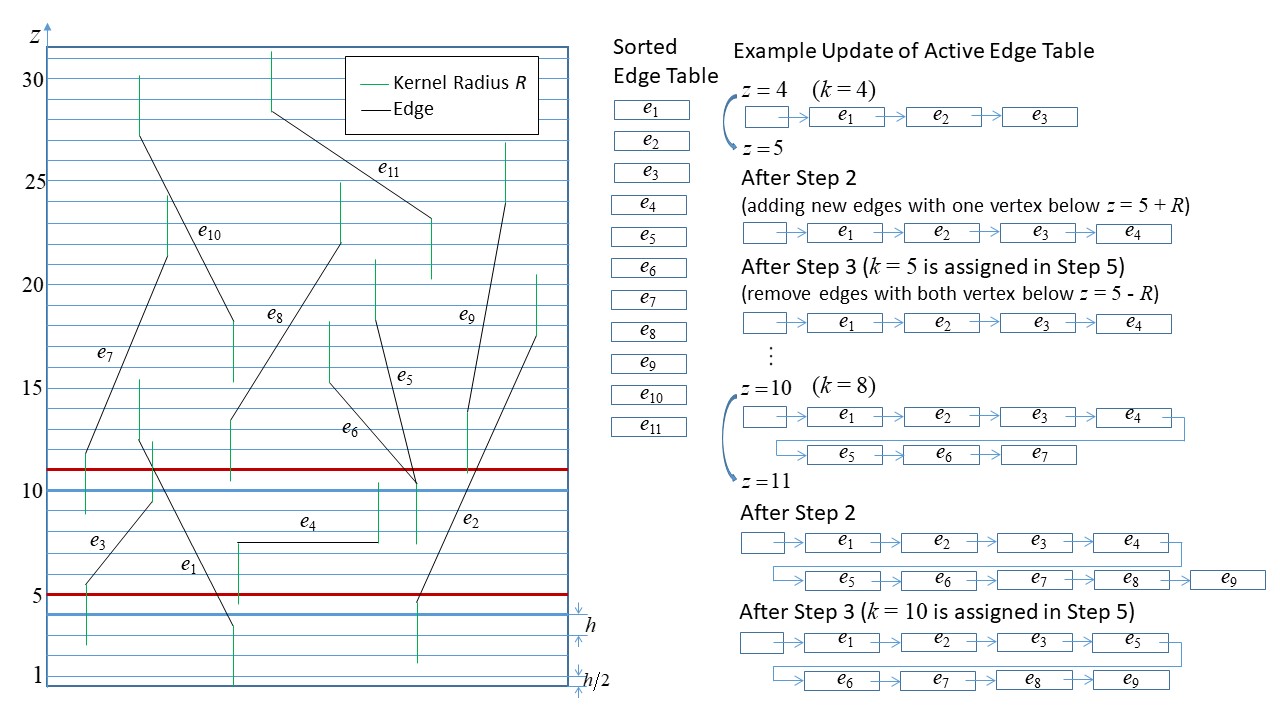}
\caption{The scanning plane algorithm can efficiently generate slides for fabricating the lattice structure represented by our method in a streaming mode -- only the skeleton edges with their swept solids intersecting with the slicing plane need to be processed.}
\label{fig:seventh}
\end{figure*}

\section{Slicing}\label{secSlicing}
In this section, we present the method for slicing the solid of a lattice structure efficiently in both memory and computational time. Benefiting from the locally supported representation of solids as formulated in Eq.(\ref{eqFuncImplicit}), only the skeleton edges with its swept sphere bounding volume \cite{larsen1999fast} intersecting the slicing plane $\mathcal{P}(z)=\{\mathbf{p} \; | \; \mathbf{p}^z=z \; (\forall \mathbf{p} \in \Re^3) \}$ will contribute to the field value of $F(\mathbf{p})$. In practice, this can be detected by a simple condition on the $z$-value of an edge's two endpoints. A set of intersected edges, denoted by $\mathcal{E}_{act}$, can be obtained as
\begin{equation}\label{eqActiveEdgeSet}
\mathcal{E}_{act}(\mathcal{P}(z))=\{ e_j=(\mathbf{v}_s, \mathbf{v}_e, r_j) \;  | \; (\mathbf{v}_s^z-R \leq \mathbf{p}^z) \cap (\mathbf{v}_e^z+R \geq \mathbf{p}^z) \},
\end{equation}
where we have \rev{$\mathbf{v}_s^z \leq \mathbf{v}_e$}{$\mathbf{v}_{s}^{z} \leq \mathbf{v}_{e}^{z}$} without loss of the generality. An algorithm in streaming mode can be used to generate slices for 3D printing a lattice structure represented by our method, which is a variant of the scanning line algorithm  \cite{hearn2004computer} and the sweeping plane algorithm \cite{Mcmains2000}. Differently, we are searching for kernels which contribute to the function value $F(\mathbf{p})$ with $\mathbf{p}^z=z$. The spherical swept volumes of convolution kernels using the support size $R$ as sphere radius are considered in our algorithm. 

Two lists of edges are constructed in our algorithm: 1) a list of all edges $\mathcal{E}$ in $\Omega$ sorted by the $z$-coordinates (in an ascending order) of their `lower' vertices and 2) a list of active edges $\mathcal{E}_{act}$ according to a slicing plane $\mathcal{P}(z)$. With the help of $\mathcal{E}_{act}$, the field-value for a point $\mathbf{q} \in \mathcal{P}(z)$ can be computed by only using the edges in $\mathcal{E}_{act}$ and using Eq.(\ref{eqFuncImplicit}). Therefore, a binary image with a user specified resolution can be generated by efficiently evaluating if $F(\mathbf{q})\leq 0$ (inside the solid) or $F(\mathbf{q}) > 0$ (outside the solid). The resultant binary image can be directly applied to the \textit{Digital Light Processing} (DLP) based 3D printer \cite{huang2014image}. Resultant binary images of example models can be found in Fig.\ref{fig:sliceimages} in Section \ref{secResult}.

When changing the slicing plane from $\mathcal{P}(z)$ to $\mathcal{P}(z+t)$ with $t$ being the layer thickness, we need to update the list of active edges. As all the edges in $\mathcal{E}$ have been sorted in an ascending order by the $z$-coordinate of their first vertex, the new list of active edges can be efficiently obtained if we still record the index of the first remaining edge in $\mathcal{E}$. We search the edges in $\mathcal{E}$ starting from this one until reaching an edge the swept solid of which is completely above $\mathcal{P}(z+t)$. Steps of our slicing algorithm are given below.
\begin{itemize}
\item \textbf{Step 1:} Initializing $k=1$, $z=\frac{t}{2}$ and  $\mathcal{E}_{act}=\emptyset$.

\item \textbf{Step 2:} Repeatedly checking edges $e_{j} \in \mathcal{E}$ with $j=k,k+1,\cdots,m$ until reaching an edge $e_{m}=(\mathbf{v}_p,\mathbf{v}_q, r_m)$ with $\mathbf{v}_p^z-R > z$ (i.e., the swept solid of $e_{m}$ is completely above $\mathcal{P}(z)$) and adding all edges $e_{j,j=k,\cdots,m-1}$ into $\mathcal{E}_{act}$.

\item \textbf{Step 3:} Checking all edges in $\mathcal{E}_{act}$ and removing an edge $e_{i}=(\mathbf{v}_s,\mathbf{v}_e)$ from $\mathcal{E}_{act}$ if $\mathbf{v}_e^z + R < z$ (i.e., its swept solid is completely below $\mathcal{P}(z)$).

\item \textbf{Step 4:} Evaluating the field value of points on the slicing plane $\mathcal{P}(z)$ by using the kernels defined on all edges in $\mathcal{E}_{act}$.

\item \textbf{Step 5:} Let $z=z+t$ and $k=m$;

\item \textbf{Step 6:} Go back to Step 2 until all planes have been slices.
\end{itemize}

The memory usage of our slicing algorithm is very limited. In summary, only the edges in $\mathcal{E}_{act}$ and the binary image for a slicing plane need to be stored in the main memory. The rest skeleton edges (i.e., $\mathcal{E}$) for a lattice structure can be processed in an out-of-core manner. \rev{}{Edges are sorted by external sorting, which also has a lightweight memeory consumption.} Detail statistics of memory consumption \rev{}{(without counting the external sorting)} for example lattice structures will be provided in Section \ref{secResult}. 

Our slicing algorithm can also be applied to generate G-code of tool-path for filament-based deposition or laser sintering. First of all, a marching square algorithm \cite{maple2003ms} is applied to a binary image to generate the rough boundary curves of a model. Then, the intersection-free smoothing and simplification algorithm \cite{huang2013slicing} can be applied to generate topology-preserved boundary curves of the model. \rev{}{The combination of the binary image and the marching square algorithm can guarantee that no degenerated case will be given in slicing, as long as the resolution of the image is large enough \cite{huang2013slicing}.} The bone model shown in Fig.\ref{fig:3DPrintedBone} is fabricated by tool-paths generated in this method on a selective laser melting based 3D metal printer.

\section{Computation for Spatially Graded Density}\label{secDensity}
\begin{figure*}
\includegraphics[width = \linewidth]{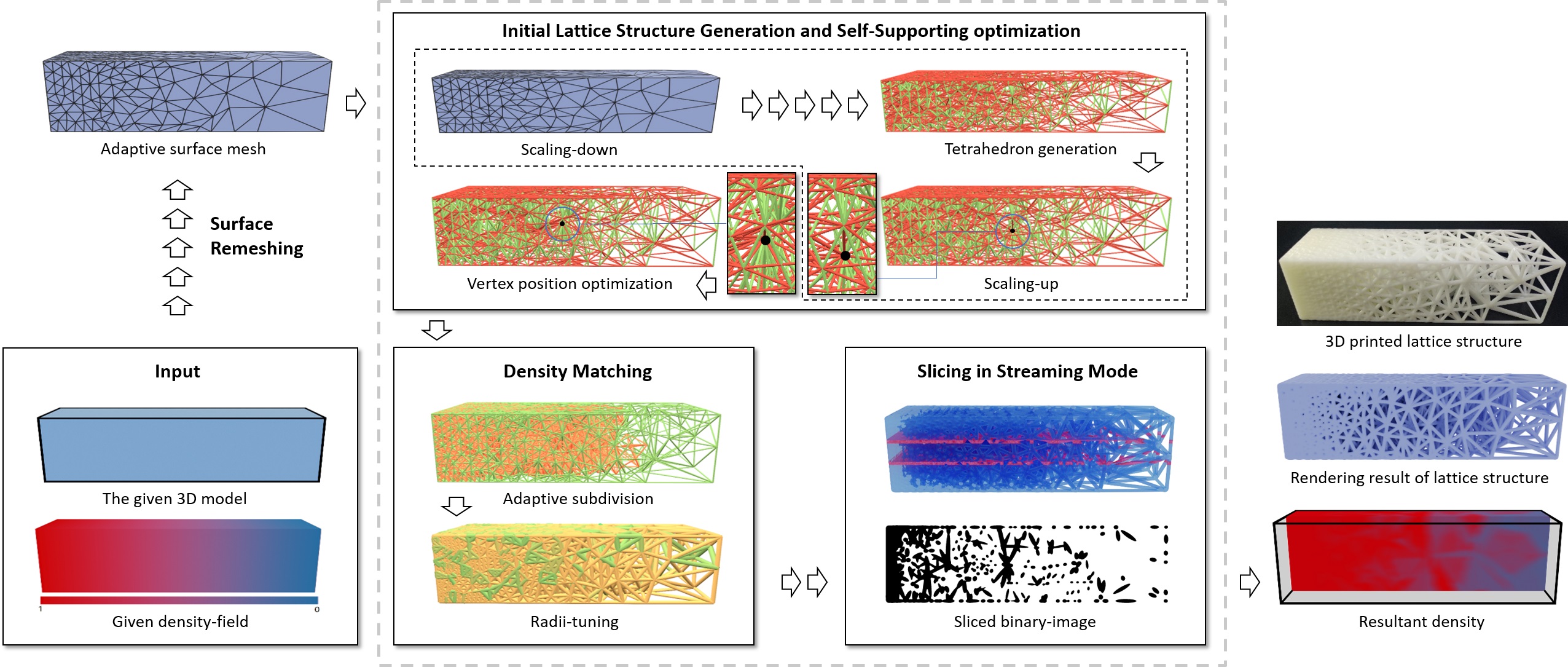}
\caption{The framework of our method to generate a lattice structure for spatial graded density consists of three major parts -- 1) adaptive surface / tetrahedral mesh generation, 2) optimization for self-supporting and 3) final density matching. By applying the slicing algorithm in streaming mode, the binary images for every slides can be generated for additive manufacturing.}
\label{fig:OverviewSpatialGradedDensity}
\end{figure*}

This section presents a framework to automatically generate a lattice structure according to an input distribution of density. Given the voxel representation $\mathcal{V}={V_{i,j,k}}$ of a given mesh model $\mathcal{M}$, the required density can be specified for each voxel $V_{i,j,k}$ as $\rho_{i,j,k}$. The problem to be solved is to construct a lattice structure $\Omega$ inside $\mathcal{M}$ so that the density of its corresponding solid $\mathcal{S}(\Omega)$ inside $V_{i,j,k}$ satisfying
\begin{equation}\label{eqDensityDesign4LatticeStructure}
\rho(\mathcal{S}(\Omega) \cap V_{i,j,k}) \approx \rho_{i,j,k}    
\end{equation}
where the density $\rho(\mathcal{S}(\Omega) \cap V_{i,j,k})$ is evaluated by the \textit{Monte Carlo integration}. Whether a sample point inside the solid $\mathcal{S}(\Omega)$ can be evaluated by the implicit function defined in Eq.(\ref{eqFuncImplicit}) efficiently.

\subsection{Overview}\label{subsecOverviewDensity}
Our framework for generating spatially graded density consists of three major parts, which are explained below with the help of illustration given in Fig.\ref{fig:OverviewSpatialGradedDensity}.
\begin{enumerate}
\item \textit{Adaptive surface / tetrahedral mesh generation:} For a given mesh model $\mathcal{M}$, we first estimate the target edge-length in different regions on its surface. An adaptive remeshing approach akin to \cite{surazhsky2003remeshing} is conducted to generate a surface adaptive mesh. Specifically, for using a material with density $\tau$ to fabricate a lattice structure with initial strut radius $r$ for realizing a target density $\rho$, the target length $\bar{L}$ of a tetrahedron can be roughly estimated as 
\begin{equation}
    \bar{L} = \max \left\{ 4r , L_{ini} \right\},
\end{equation}
where $L_{ini}$ is a solution of the density estimate formula Eq.(\ref{eqMeshEdgeLengthEstimation}) that is closest to the average edge length of the initial given mesh.
Detail calculation can be found in Appendix I. The target lengths at different surface regions are computed by the above equation to control the result of surface remeshing. After that, a tetrahedral mesh generation method \cite{si2015tetgen} is applied to construct the volumetric mesh adaptive to the surface mesh. Note that, the step of surface remeshing is very important because it is difficult to generate a locally coarse tetrahedral mesh if the surface mesh is dense. After this step, edges of the tetrahedral mesh are employed to generate the initial lattice structure $\Omega$.

\item \textit{Optimization for self-supporting:} When being fabricated by additive manufacturing, supporting structures need to be added below the large overhangs in $\Omega$. This will change density on the finally fabricated models. Therefore, we develop an algorithm to improve the self-supporting of edges in $\Omega$, which consists of two steps -- the scaling / re-scaling and the vertex re-positioning. Details are given in Section \ref{subsecSelfSupportingOptm}.

\item \textit{Density matching:} After optimizing the self-supporting of a lattice structure, its density is further adjusted in the final phase of our framework to match the required density distribution. This is implemented by applying local subdivision on tetrahedra and local adjustment for the radii of struts. Details are given in Section \ref{subsecDensityOptm}.
\end{enumerate}
With the help of this framework, we enable the inverse design of spatially graded density by using lattice structures.

\subsection{Optimization for Self-Supporting}
\label{subsecSelfSupportingOptm}
For additive manufacturing, supporting structures need to be added below the region with large overhang, which prolongs the printing process and wastes more material. More seriously, the supporting structures are hard to remove, and keeping these supporting structures will change the density distribution of a designed lattice structure. Therefore, it is important to reduce the demand of support by optimizing a design.

For all edges in a lattice structure $\Omega=(\mathcal{V},\mathcal{E})$, we define a metric of self-supporting based on the projected area of risky regions that are facing down. For an edge $e_j=(\mathbf{v}_s,\mathbf{v}_e,r_j) \in \mathcal{E}$, whether it is fully self-supported depends on the angle $\theta(e_j)$ between it and the printing direction $\mathbf{t}_P$ as
\begin{equation}\label{supportAngle}
\theta(e_j) = \arccos \left| \frac{\mathbf{v}_{e}-\mathbf{v}_{s}}{\| \mathbf{v}_{e}-\mathbf{v}_{s} \|} \cdot \mathbf{t}_P \right|
\end{equation}
with $\mathbf{t}_P$ being a unit vector. For an edge satisfying $\theta(e_j) \leq \alpha$, the strut generated by this edge is fully self-supported. Here the angle $\alpha$ is the \textit{self-supporting angle} that depends on the type of AM process and the materials used. In the rest of our paper, we conduct a widely used parameter $\alpha=\frac{\pi}{4}$. The set of fully self-supported edges is denoted as $\mathcal{E}_{S}$.

When $\theta(e_j) > \alpha$, a portion of the facing-down surface on the strut needs to be fabricated by adding supporting structures. For a cylindrical strut with skeleton as $e_j$, the projected facing down area is $A(e_j)= 2 r_j L(e_j)$ with $L(e_j)$ being the length of $e_j$. According to our analysis on a cylinder, the percentage of projected area that needs to add support is a function $g(\theta)$ in terms of the angle $\theta$ between the cylinder's axis and the printing direction. 
Detail analysis can be found in Appendix II. For the sake of computational simplicity, we conduct a polynomial to approximate $g(\theta)$. When $\alpha = \pi /4$, it is
\begin{equation}
\label{eqRiskyPercentage}
g(\theta) \approx \left\{
\begin{aligned}
\Sigma_{i=0}^5 a_{i} \theta^i & & (\theta>\pi/4)\\
0 & & (\theta \leq \pi/4)
\end{aligned}
\right.
\end{equation}
where $a_i=[-0.02,-0.31,1.44,-1.11,0.58,-0.16]$.


Based on this analysis, we define a metric of self-supporting for the lattice model as
\begin{equation}\label{eqSupportMetric}
    \Gamma(\Omega) = \frac{\sum_{e_j \in (\mathcal{E} \setminus \mathcal{E}_S)} r_j L(e_j) g(\theta_j) }{
    \sum_{e_j \in \mathcal{E}} r_j L(e_j) }
\end{equation}
which is the percentage of projected areas that needs additional support; the smaller the better. Besides of $\Gamma(\Omega)$, we also define a length percentage of completely self-supported struts as 
\begin{equation}\label{eqSupportLengthMetric}
    \Psi(\Omega) = \frac{\sum_{e_j \in \mathcal{E}_S} L(e_j)}{
    \sum_{e_j \in \mathcal{E}} L(e_j) } \times 100\%
\end{equation}
which is the higher the better.

\begin{figure}[t]
\centering
\subfigure[Without optm.
]{
\includegraphics[width=0.28\linewidth]{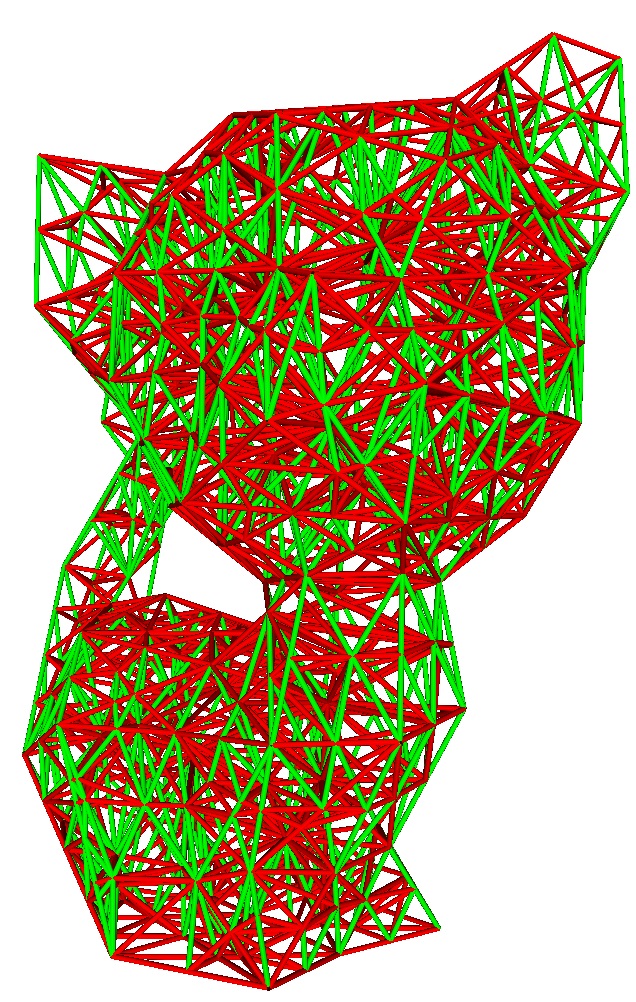}
}
\subfigure[After SO
]{
\includegraphics[width=0.30\linewidth]{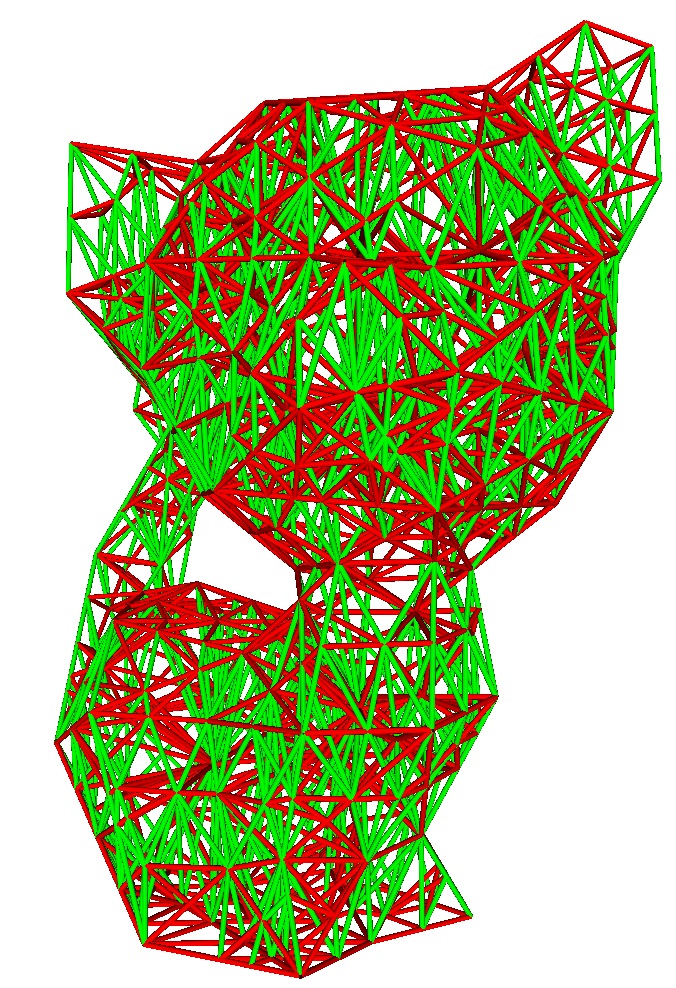}
}
\subfigure[After PO
]{
\includegraphics[width=0.30\linewidth]{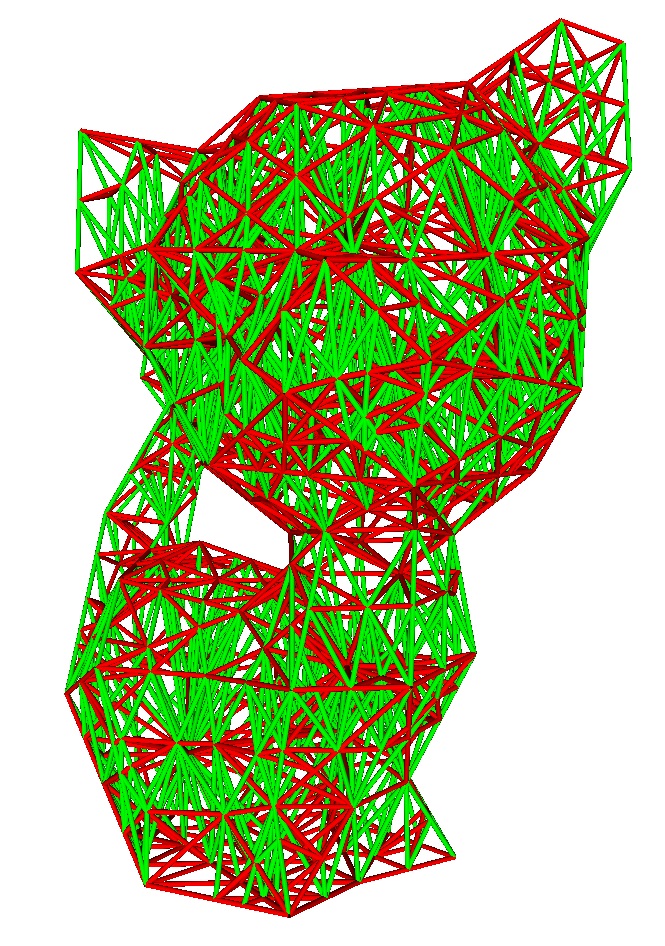}
}
\caption{An illustration for demonstrating the effectiveness of our algorithm for enlarging the percentage of self-supported edges, where edges need to add support are displayed in red color. Two schemes, the Scaling Operation (SO) and the vertex Position Optimization (PO), are applied to the lattice structure $\Omega$ for a kitten model. The length percentages of completely self-supported edges are (a) $\Psi=27.97\%$, (b) $\Psi=41.67\%$ and (c) $\Psi=49.38\%$ respectively. 
When the same radius is employed for all struts, the metric of self-supporting can be significantly reduced from (a) $\Gamma=0.6275$ to (b) $\Gamma=0.4830$, and then to (c) $\Gamma=0.3726$.
}
\label{fig:SelfSupportingOptm}
\end{figure}

We develop two schemes to improve the value of $\Gamma(\Omega)$ on a lattice structure $\Omega$: \textit{Scaling Operation} (SO) and vertex \textit{Position Optimization} (PO). As illustrated by the kitten model shown in Fig.\ref{fig:SelfSupportingOptm}, these schemes can
effectively reduce the percentage of projected areas that need to add support in a lattice structure.

\begin{figure}[t]
\centering
\includegraphics[width=\linewidth]{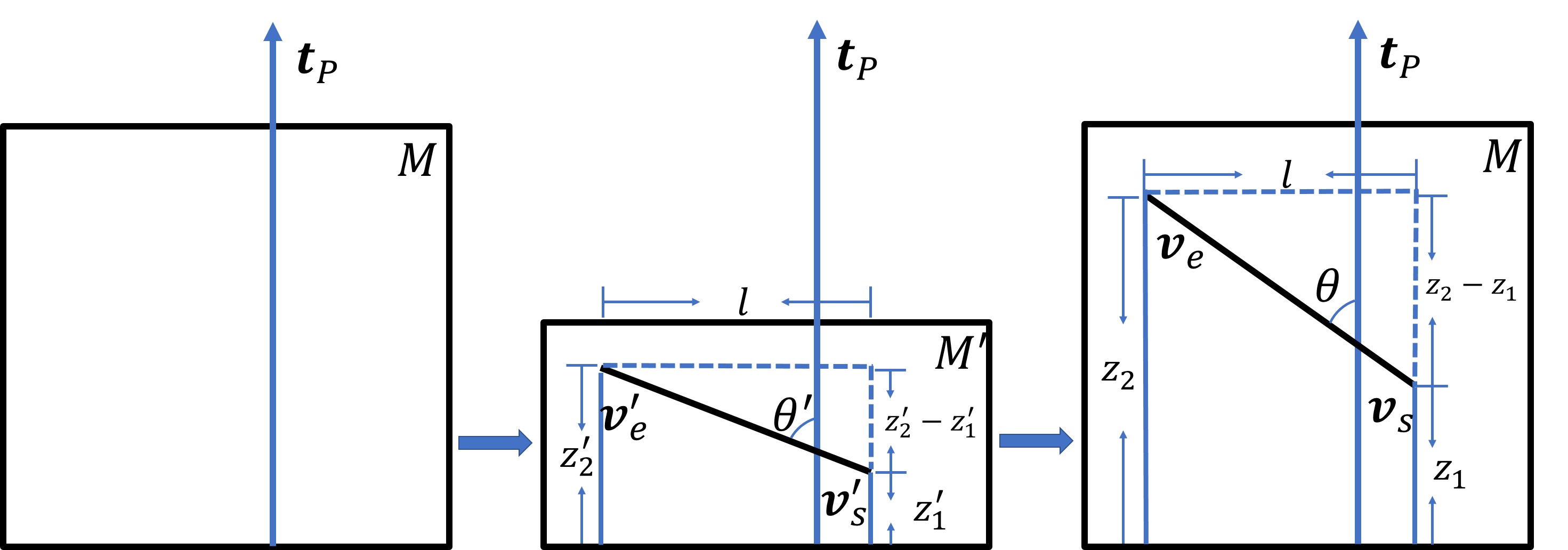}	
\caption{The edges with large overhang generated in a scaled model (with height collapsed along the printing direction $\mathbf{t}_P$ -- as shown in the middle) have good chance to become self-supported after being scaled back to the model's original height. 
}\label{fig:scaling4SelfSupporting}
\end{figure}

\subsubsection{Scaling}
For a given 3D printing direction $\mathbf{t}_P$ and an edge $e_j=(\mathbf{v}_s,\mathbf{v}_e)$ bounded in a box $M$ as shown in Fig.\ref{fig:scaling4SelfSupporting}, the angle $\theta$ between $\mathbf{t}_P$ and $\mathbf{v}_s \mathbf{v}_e$ will become large when they are not perpendicular. Without loss of the generality, we can assume $\mathbf{t}_P=(0,0,1)$, $\mathbf{v}_s=(x_1,y_1,z_1)$, $\mathbf{v}_e=(x_2,y_2,z_2)$ and the scaling factor as $k$. After scaling, the new positions of the edge become $\mathbf{v}'_s=(x_1,y_1,z'_1)$ and $\mathbf{v}'_e=(x_2,y_2,z'_2)$ with $z'_1=z_1/k$ and $z'_2=z_2/k$. Then, we get
\begin{equation}\label{scalingAngle}
\centering
\begin{aligned}
\theta\left(\mathbf{v}_{s} \mathbf{v}_{e}\right)=\frac{\pi}{2}-\arctan \left(\frac{\left|z_{2}-z_{1}\right|}{l}\right),\\ \theta\left(\mathbf{v}'_{s} \mathbf{v}'_{e}\right)=\frac{\pi}{2}-\arctan \left(\frac{\left|z_{2}-z_{1}\right|}{k l}\right),
\end{aligned}
\end{equation}
where 
$l=\sqrt{\left(x_{1}-x_{2}\right)^{2}+\left(y_{1}-y_{2}\right)^{2}}$. 
It is easy to find $\theta\left(\mathbf{v}_{s} \mathbf{v}_{e}\right) < \theta\left(\mathbf{v}'_{s} \mathbf{v}'_{e}\right)$ when $k>1$.

\begin{figure*}[t]
\centering
\includegraphics[width=\linewidth]{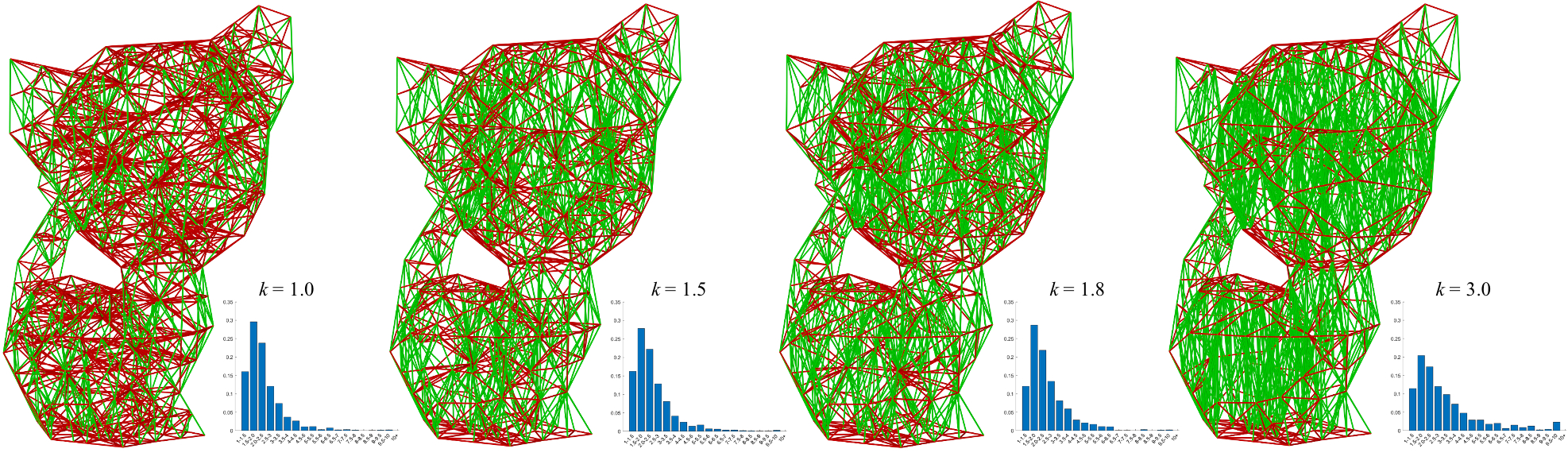}	
\caption{A study to find the balance between the level of self-supporting (can be enhanced by using large $k$) and the quality of constructed tetrahedral mesh (will be reduced by using large $k$). 
From left to right, the length percentage of completely self-supported edges $\Psi$ are $27.97\%$, $41.67\%$, $52.38\%$ and $79.28\%$ respectively (from left to right). 
When the same radius is used for all struts, the metric of self-supporting $\Gamma$ gives the values of $0.6275$, $0.4230$, $0.3316$ and $0.1271$ (from left to right).}\label{fig:k-aspectratio}
\end{figure*}

Based on this analysis, we can improve the self-supporting of a lattice structure during its construction by scaling. Specifically, we first compress the space for constructing the lattice structure by a factor ${1}/{k}$ along the printing direction $\mathbf{t}_P$. After constructing the tetrahedral mesh as an initial lattice structure in the compressed space, we re-scale the mesh back to the original size by a factor $k$ along the direction $\mathbf{t}_P$. Lattice structures constructed with the help of this scaling step have more self-supported edges (see Fig.\ref{fig:SelfSupportingOptm}(a) and \ref{fig:SelfSupportingOptm}(b) for an example). 
Using $k$ with too large value will lead to very sparse vertices generated inside a model. Moreover, the quality of tetrahedra can be very poor when using a large $k$, which can be quantitatively evaluated by the aspect ratio on all tetrahedra. In short, the aspect ratio of a tetrahedron $T_i$ is calculated by $R(T_i)={h_{\max}}/{h_{\min}}$, where $h_{\max}$ and $h_{\min}$ are the maximum and the minimum distances from a vertex to its opposite face inside $T_i$. The ideal value of $R(T_i)$ is $1.0$ and the quality of a mesh is considered as poor when many tetrahedra have the aspect ratio greater than $5.0$. The histograms of aspect ratios for tetrahedral meshes generated by using different $k$ are studied to find a good balance between the quality of mesh and the level of self-supporting (see Fig.\ref{fig:k-aspectratio}). According to this study, we usually employ $k \in [1.5, 1.8]$ in practice. 

\subsubsection{Vertex re-positioning}
When moving a vertex $\mathbf{v}$, all edges linked to it (denoted the set as $\mathcal{E}_{\mathbf{v}}$) will be changed. Therefore, we can potentially move the vertices to generate more fully self-supported edges. We define an objective function to evaluate the self-supporting property around a vertex $\mathbf{v}$ as
\begin{equation}\label{eqLocalSelfSupportingF}
J(\mathbf{v})= \frac{
\sum_{e_{j} \in \mathcal{E}_{\mathbf{v}}}
{ r_i L(e_{j}) g(\theta_i)}
}{
\sum_{e_{i} \in \mathcal{E}_{\mathbf{v}}}
{ r_i L(e_{i})
}
}
\end{equation}
which need to be minimized. Moreover, in order to prevent intersections between tetrahedra, the new position of $\mathbf{v}$ and $\bar{\mathbf{v}}$ will be confined in a limited space. The optimization is formulated as
\begin{equation}\label{eqOptiPosProblem}
\centering
\begin{aligned}
\bar{\mathbf{v}} = \arg \min_{\mathbf{v}} J(\mathbf{v})\\ 
s.t. \; \|\mathbf{v}-\mathbf{o}\| \leq \tau
\end{aligned}
\end{equation}
where $\mathbf{o}$ and $\tau$ are the center and the radius of an inscribed sphere of the polyhedron formed by the vertices incident to the vertex $\mathbf{v}$. \rev{}{When solving the above optimization problem, the movement of vertex is conducted along the gradient direction. Together with the step length in every iteration, it determines how the vertices should be moved to improve the self-supporting property of the lattice.}


\begin{figure}[t]
\centering
\subfigure[]{
\includegraphics[width=.5\linewidth]{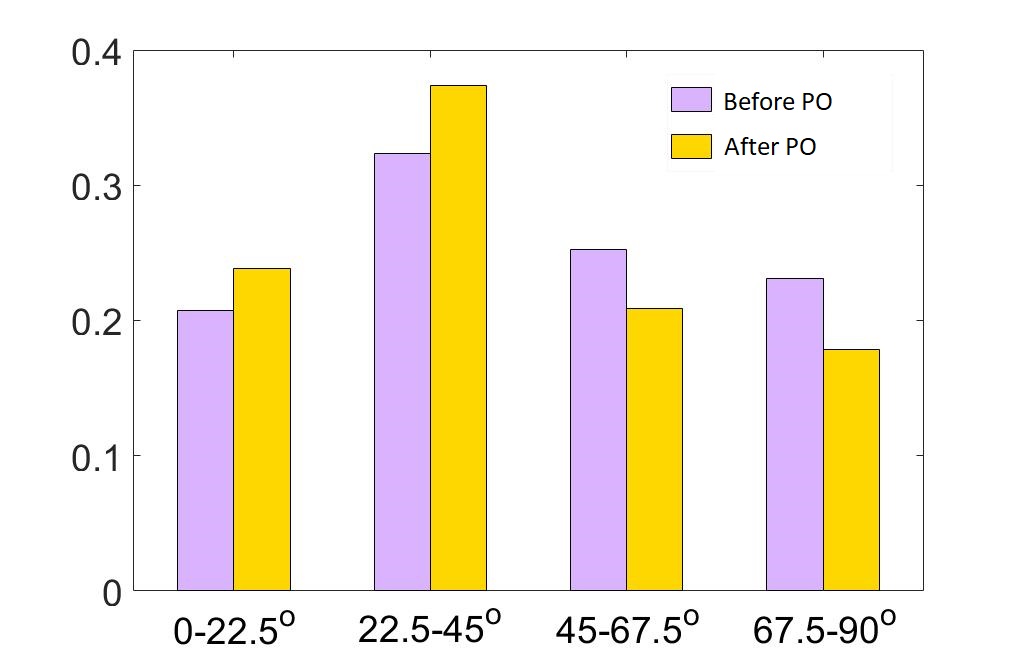}
}
\subfigure[]{
\includegraphics[width=.95\linewidth]{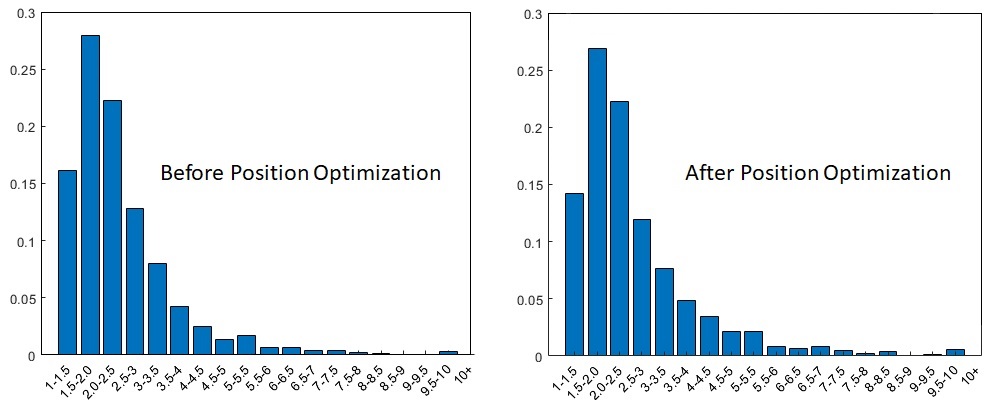}
}
\caption{Statistic visualization for (a) the histogram of angles between edges and the printing direction and (b) the aspect ratio of tetrahedra before vs. after position optimization.} \label{fig:VerPosOptmRes}
\end{figure}

The optimization is randomly applied to all the interior vertices one by one. 
The iteration stops when 
1) no more vertex can move, 
2) no more vertex's movement can reduce the value of $\Gamma(\mathbf{v})$ 
or 3) the maximum number of iterations (i.e., $100$ used in our implementation) have been reached. 
Figure \ref{fig:VerPosOptmRes}(a) gives a histogram chart to show the distribution of angles between edges and the printing direction $\mathbf{t}_P$, where the vertical axis gives the percentage of edges in terms of length. Moreover, it is interesting to study the aspect ratios of tetrahedra before and after position optimization. As shown in Fig.\ref{fig:VerPosOptmRes}(b), the distributions of aspect ratios do not change too much, which is benefit by constraining the magnitude of movement in Eq.(\ref{eqOptiPosProblem}). 

\subsection{Density Matching}\label{subsecDensityOptm}
To explore the porosity of lattice structure, we initially define the radius $r_i = 2r_{\min}$ on all edges $\{e_i\}$ with $r_{\min}$ being the smallest feature size that can be reliably fabricated on a 3D printer. The radii will be adjusted for density matching below. 

\begin{figure}[t]
\centering
\includegraphics[width=\linewidth]{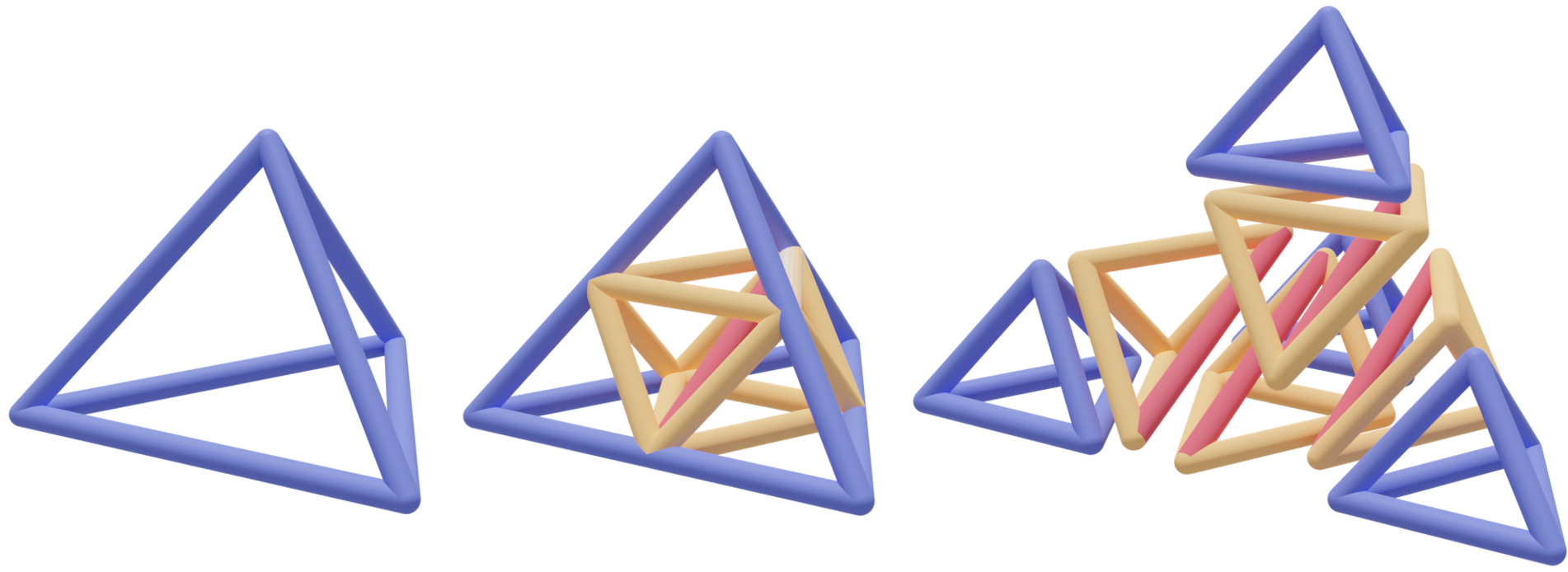}
\caption{The subdivision operator for density matching -- the density can be increased by more than doubled after subdivision. Most of the newly generated edges are support-free, except for the red one\rev{}{~(which is kept in the structure)}. 
}\label{fig:TetSubd}
\end{figure}

To satisfy the density distribution that is already defined as input of voxel-based function values (i.e., Eq.(\ref{eqDensityDesign4LatticeStructure})), two operators are developed to adjust the density given on a lattice structure. 
\begin{enumerate}
\item \textbf{Subdivision:} When the density of material inside a tetrahedron (formed by the struts on its 6 edges) is not large enough, we would subdivide a tetrahedron $T_i$ into 8 tetrahedra by splitting in the middle of each edge (as illustrated in Fig.\ref{fig:TetSubd}). From the geometry of tetrahedra obtained after subdivision, it can be observed that the newly created edges (except the red one in Fig.\ref{fig:TetSubd}) are always parallel to one of the 6 edges on $T_i$. As a result, the responding edges will be self-supported if the original edges of $T_i$ are. The total length of edges is increased from $\sum_{i=1}^6 l_i$ (with $l_i$s being the length of a tetrahedron's six edges) to $l_r+2\sum_{i=1}^6 l_i$ with $l_r$ being length of the red edge as shown in Fig.\ref{fig:TetSubd}. Therefore, the density in the volume of the original tetrahedron is more than doubled when using the same radii for the struts of newly generated edges. 

\item \textbf{Radii-Tuning:} When the density of material inside a tetrahedron $T_i$ is larger than a target value, we match the designed density by reducing the value of $r_k$ on every edge $e_k \in T_i$. Note that, when changing $r_k$, the radius of strut on $e_k$ is changed monotonically but not linearly (see Fig.\ref{fig:FieldComp}(b) for an illustration). Note that the minimal radii of edges should be bounded by the smallest feature size that can be fabricated by 3D printing. 
\end{enumerate}
These two operators are applied in our algorithm to generate a lattice structure matching the desired density-distribution.

Our algorithm generates the density distribution of a lattice structure according to the input in four steps:
\begin{enumerate}
\item First, we assign the desired density $\bar{\rho}_{T}$ to every tetrahedron $T$ by the target density given to voxels in $T$ -- the maximal one is employed when $T_i$ contains more than one voxels. Here $V_{i,j,k}$ is considered as being contained by $T$ when its center is inside $T$.  

\item For each tetrahedron $T$, we denote its current density as $\rho_{T}$ and the its density after making all edges' radii doubled as $\rho_{T}^*$. When $\rho_{T}^*<\bar{\rho}_{T}$, we subdivide this tetrahedron into 8 tetrahedra recursively until this condition is satisfied in every new tetrahedron. For a small tetrahedron does not contain the center of any given voxel, the desired density is evaluated at the center of this tetrahedron by linearly interpolating the densities given at the centers of voxels.

\item For each tetrahedron $T$, if it satisfies $\rho_{T}<\bar{\rho}_{T}<\rho_{T}^*$, we amplify the radii of edges on this tetrahedron by using the binary searching strategy to match the target density $\bar{\rho}_{T}$.

\item Lastly, we check the scaled radii of all edges in every voxels to generate a more accurate density matching. Specifically, for every voxel $V_{i,j,k}$, we try to determine a common radii-scaling ratio for all edges intersected with this voxel so that the density $\rho_{i,j,k}$ is matched more accurately. If the determined scaling ratio is less than the current ratio stored on an edge, the ratio is updated by the newly determined one. After checking all voxels, the struts of all edges are scaled by the most updated ratios. 
\end{enumerate}
After applying these fours steps of our algorithm, the material density given on the lattice structure can well match the desired density distribution. 

\section{Results and Discussion}\label{secResult}
The approach described in the previous sections has been implemented using C++, based on a 4.00 GHz Intel Core i7-4790K and 16GB memory. Using this implementation, a variety of case studies and comparisons will be presented in this section to validate the approach.

\begin{table}[t]
\centering 
\caption{Statistic of Triangular Meshes Generated by MC}\label{tab:StatisticsMC}\vspace{8pt}
\small
\begin{tabular}{l|l|l|l|l}
  \hline
  \textbf{Models} & {Bunny} & {Bone} & {Finger} & {Kitter-HR}\\
  \# of struts & 123,610 & 75,484 & 359,212 & 14,063,027  \\
\hline\hline
  Approx. Err. &  \multicolumn{4}{c}{$5\%$ of strut radius $r = 0.5 mm$} \\
  \hline
  Box Size & \multicolumn{4}{c}{$ 0.0847~mm \times 0.0847~mm \times 0.0847~mm $} \\
  \hline
                   & $1368$ & $1384$ & $904$ & $2864$ \\
  Res. of MC & $\times 1368$ & $\times 1384$ & $\times 904$ & $\times 2864$ \\
                   & $\times 1447$ & $\times 1826$ & $\times 1758$ & $\times 2788$ \\
  \hline
  \hline
  \# of Triangles & 355.7M & 304.9M & 226.2M & 2,359M  \\
  \hline
\end{tabular}
\end{table}

\subsection{Memory-Efficient Representation}\label{subsecResultMemory}
Traditionally, lattice structures are represented as triangular meshes to allow easy integration into the 3D printing pipeline \cite{gao2015status}. To convert an implicitly represented lattice structure (e.g., the scheme used in this work) to a triangular mesh, the \textit{Marching Cubes} (MC) method \cite{lorensen1987marching} is often used. However, for large-scale lattice structures, the MC method produces a huge number of triangles. As can be found from Table \ref{tab:StatisticsMC}, 226M to 2,359M triangles were generated for the models tested when setting the approximation error as $5\%$ of the strut radii during triangulation.

Recently, a new method named \textit{Lattice Structure Lightweight Triangulation} (LSLT) was proposed to triangulate lattice structures \cite{chougrani2017lattice}. This method can reduce the number of the generated triangles, as shown by the statistics in Table \ref{tab:StatisticsFileSizeComparison}. Nevertheless, it still generated too many triangles (26M to 258M) for the tested models when the number of struts increases significantly. As a result, the memory consumption of this method is still too large for slicing and toolpath planning algorithms to run properly.

\begin{table}[t]
\centering
\caption{Comparisons of float numbers (or file sizes) for different algorithms.}\label{tab:StatisticsFileSizeComparison}\vspace{8pt}
\small
\begin{tabular}{l|l|l|l|l}
  \hline
  \textbf{Models} & Bunny & Bone & Finger & Kitty-HR\\
  \# of struts & 123,610 & 75,484 & 359,212 & 14,063,027  \\
  \hline
  \hline
        & \multicolumn{4}{c}{\# of Triangles$^{\dag}$ (Unit: $10^6$)}   \\
  \hline
  MC & 355.7 & 304.9  & 226.2 & 2358.6   \\
  LSLT  & 26.91  & 19.69  & 27.83  & 258.7  \\
  \hline
  \hline
        & \multicolumn{4}{c}{File Size (Unit: MB)}   \\
  \hline
  MC    & 12551.2 & 10759.1 & 789.17 & 83225.4   \\
  LSLT  & 949.2  & 695.1  & 980.9  & 9128.5    \\
  \hline 
  \textbf{Ours}  & 3.36  & 0.76  & 10.7  & 475.63  \\
  \hline
\end{tabular}
\begin{flushleft}
$^{\dag}$Note that, the number of triangles generated by LSLT are estimated according to their ratios to the result of MC, which are provided by \cite{chougrani2017lattice}.
\end{flushleft}
\end{table}

By contrast, the proposed method can significantly reduce the memory consumption (see the File Size block given in Table \ref{tab:StatisticsFileSizeComparison}). This is essentially achieved by completely dropping the explicit triangular mesh representation scheme, and instead, by representing lattice structures implicitly using convolution surfaces with line segments as skeletons. The corresponding solid models of the lattice structures are generated on-site in a streaming manner. Table \ref{tab:StatisticsFileSizeComparison} shows the comparison of the file size for storing the same lattice structures with the MC method, the LSLT method, and our method. Clearly, the proposed method resulted in files with much lighter size for representing lattice structures.

\subsection{Density Matching}\label{subsecResultDensityMatching}
Having shown the memory-efficiency feature of the proposed approach, we now move to the effectiveness of our approach: generating lattices structures that match the prescribed density distribution.

\begin{figure}[t]
\centering
\subfigure{
\includegraphics[width=0.19\linewidth]{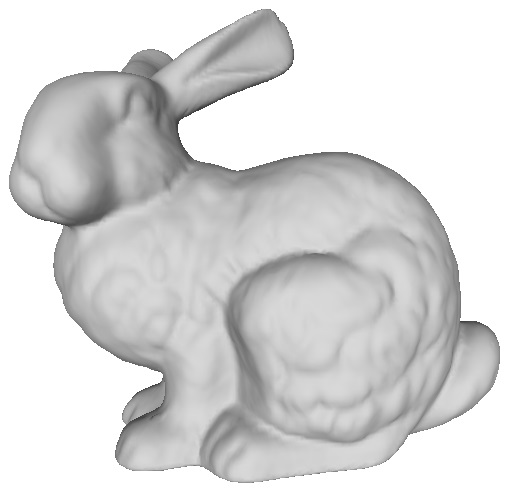}
}
\subfigure{
\includegraphics[width=0.25\linewidth]{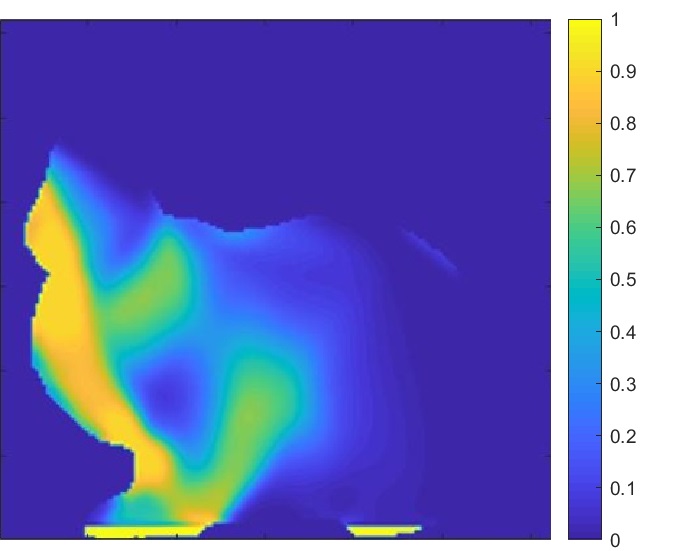}
}
\subfigure{
\includegraphics[width=0.25\linewidth]{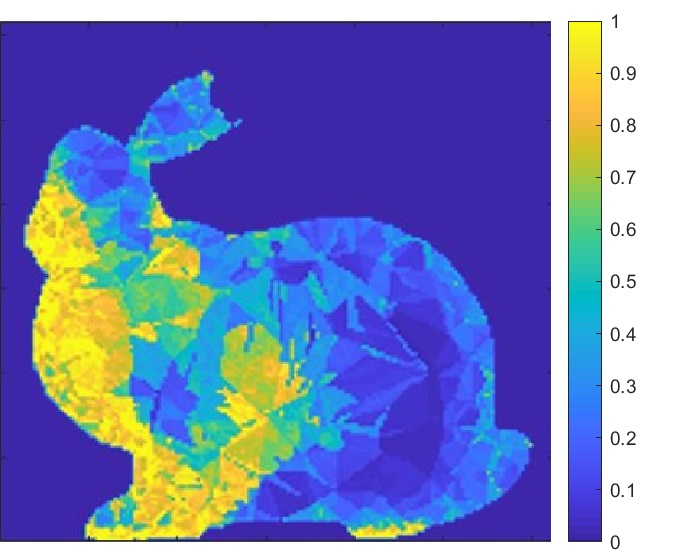}
}
\subfigure{
\includegraphics[width=0.19\linewidth]{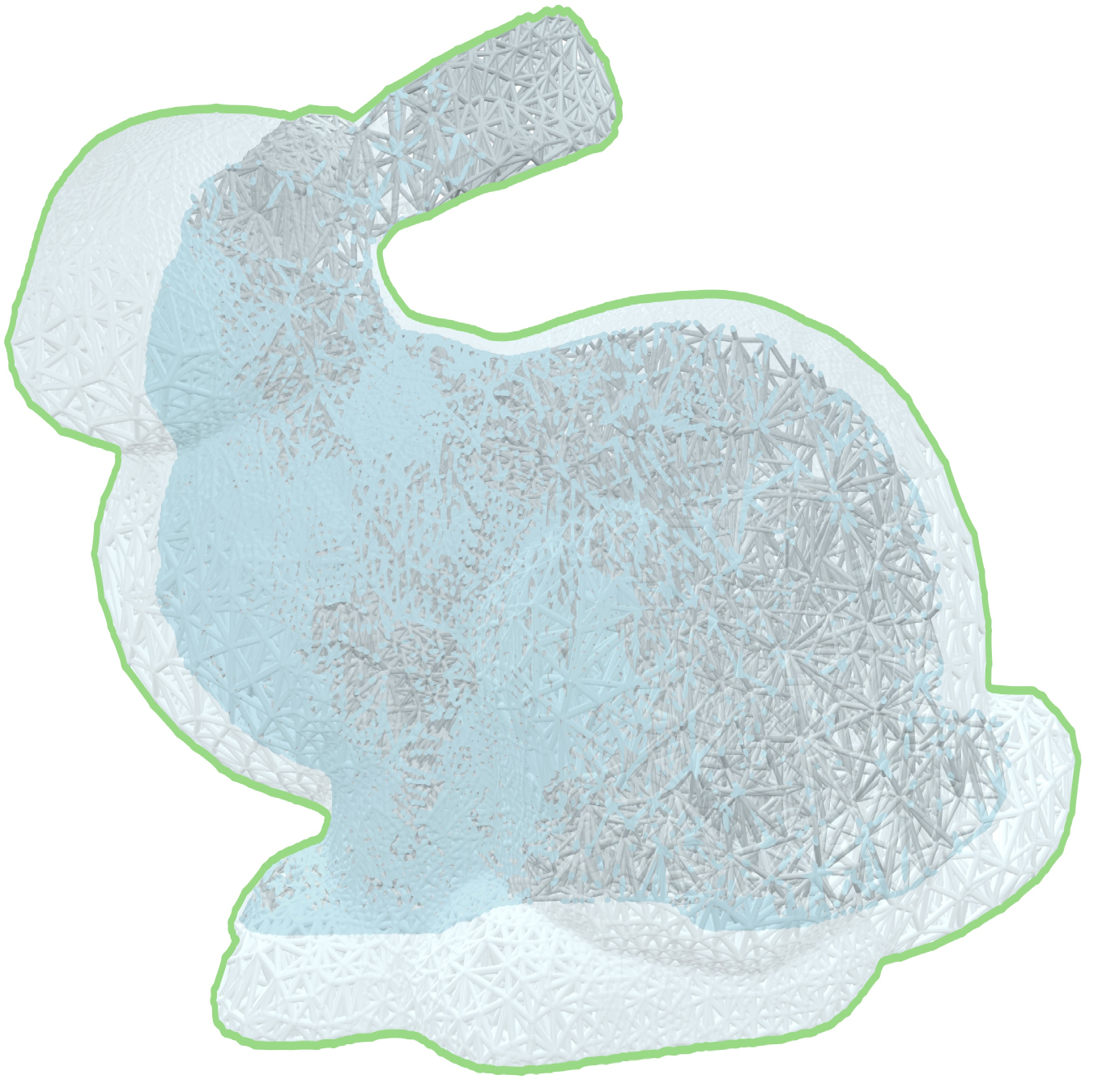}
}
\subfigure{
\includegraphics[width=0.19\linewidth]{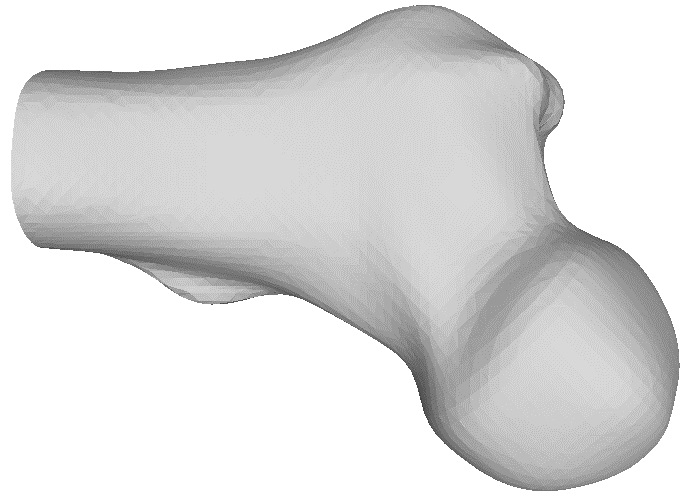}
}
\subfigure{
\includegraphics[width=0.25\linewidth]{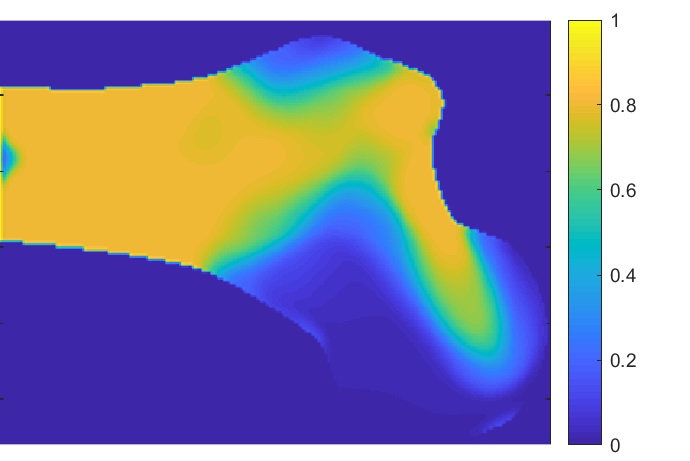}
}
\subfigure{
\includegraphics[width=0.25\linewidth]{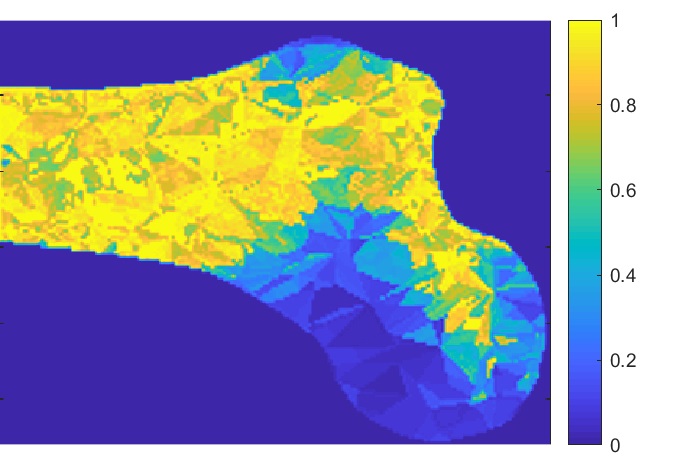}
}
\subfigure{
\includegraphics[width=0.19\linewidth]{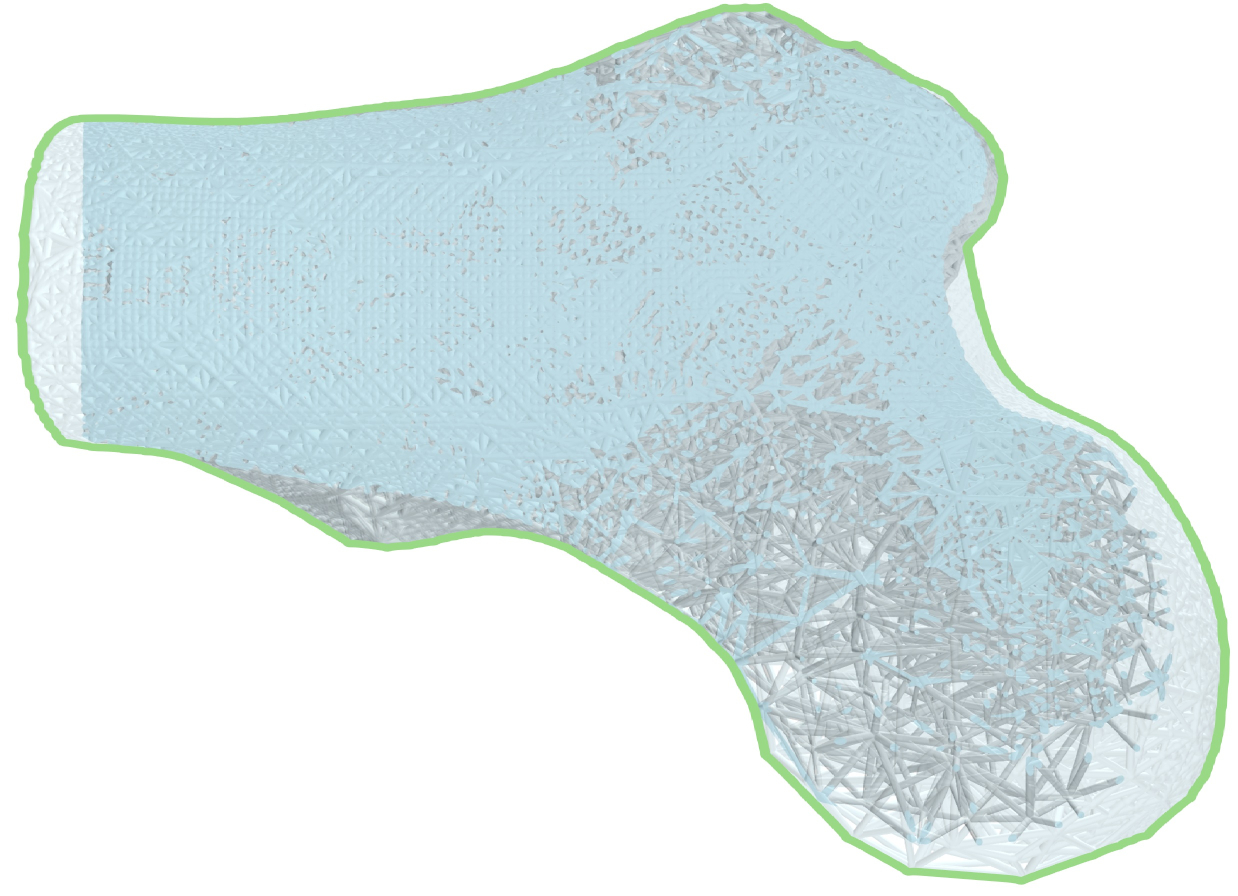}
}
\addtocounter{subfigure}{-8}

\subfigure[]{
\label{fig:DensityMatchingRes:a}
\includegraphics[width=0.19\linewidth]{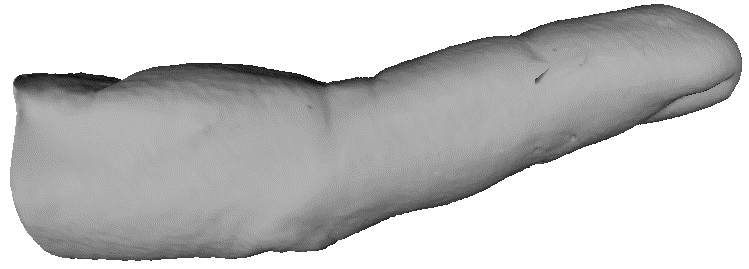}
}
\subfigure[]{
\label{fig:DensityMatchingRes:b}
\includegraphics[width=0.25\linewidth]{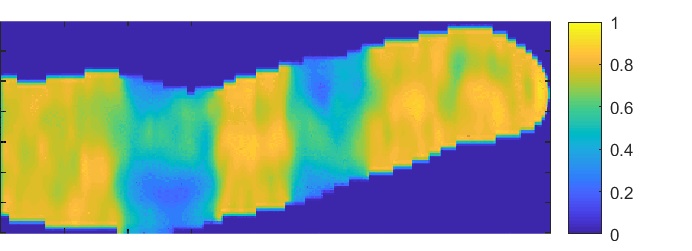}
}
\subfigure[]{
\label{fig:DensityMatchingRes:c}
\includegraphics[width=0.25\linewidth]{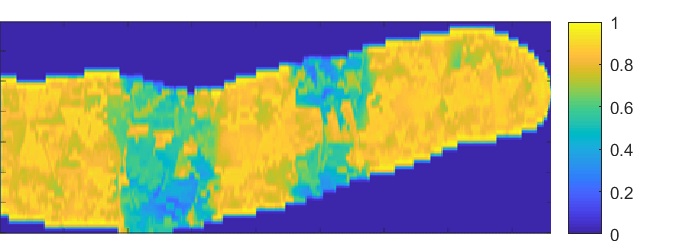}
}
\subfigure[]{
\label{fig:DensityMatchingRes:d}
\includegraphics[width=0.19\linewidth]{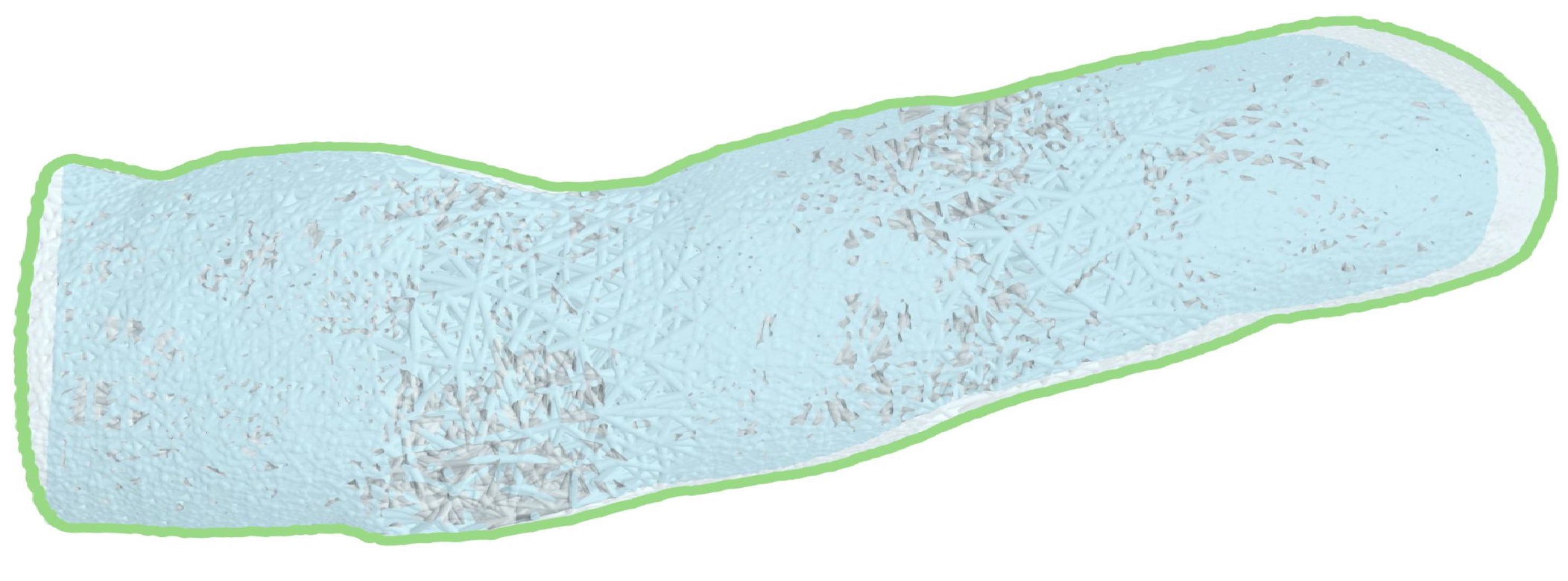}
}
\caption{Three examples to demonstrate the function of our approach in generating lattice structures for density matching: (a) the input 3D models, (b) the required density, (c) the resultant density and (d) the resultant lattice structures. Note that, the resultant lattice structures are rendered by directly applying ray-tracing in POV Ray \cite{plachetka1998pov} (i.e., no mesh surface is generated).
}
\label{fig:DensityMatchingRes}
\end{figure}

Three models were tested, and the results are given in Fig.\ref{fig:DensityMatchingRes}. The given density distribution took the form of a voxel set, and each voxel's density value was evaluated by applying the Monte-Carlo integration to the implicit solid models generated by our approach. The corresponding implicit solid models are depicted in Fig.\ref{fig:DensityMatchingRes:d}. From Fig.\ref{fig:DensityMatchingRes:c}, our method is found to generate lattice structures that match the prescribed density (Fig.\ref{fig:DensityMatchingRes:b}) very well. 

\subsection{Self-supporting Optimization}
This sub-section shows the effectiveness of the self-supporting optimization module of our approach. We compared the raw results from the \textit{tetrahedral mesh generation} (Tet-Gen) method \cite{si2015tetgen} with those optimized by ours (i.e., with two additional optimization steps: SO and PO). 

\begin{figure}[t]
\centering
\subfigure[Lattice structures directly generated by Tet-Gen \cite{si2015tetgen} -- the values of $\Psi$ are $30.4\%$ (Teddy), $28.0\%$ (Kitten), $27.7\%$ (Camel) and $29.9\%$ (Horse)]{
\includegraphics[width=0.17\linewidth]{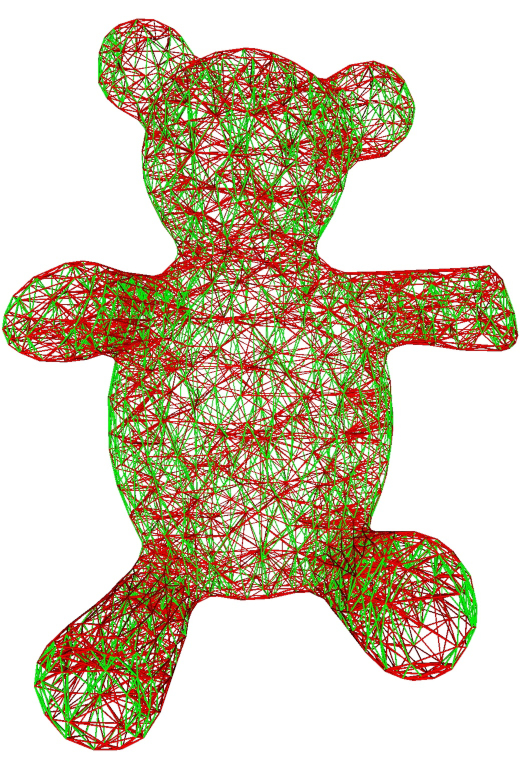}
\includegraphics[width=0.16\linewidth]{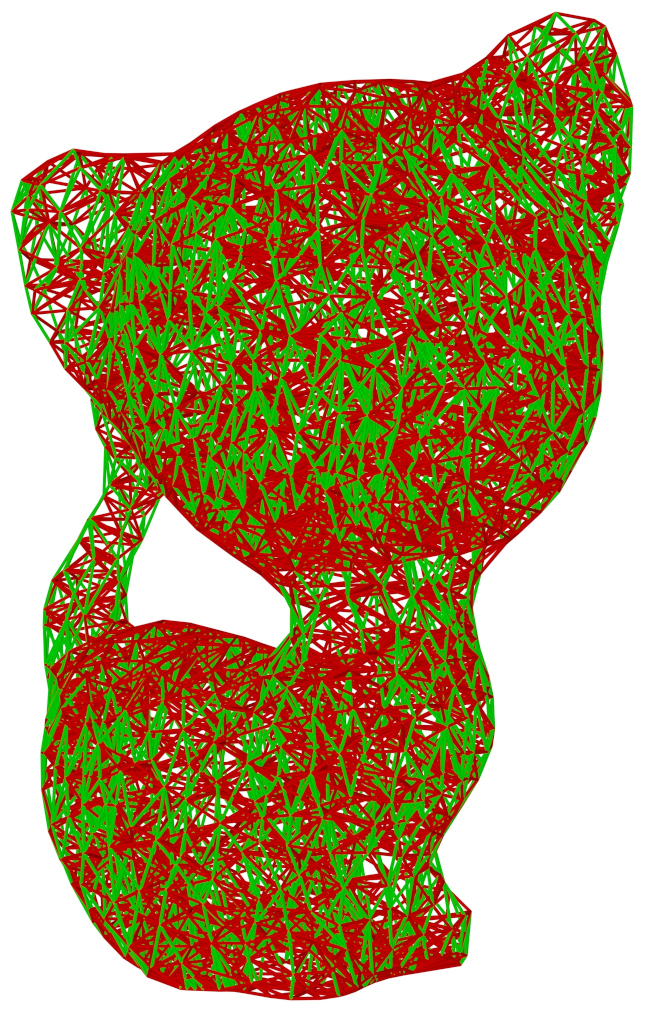}
\includegraphics[width=0.25\linewidth]{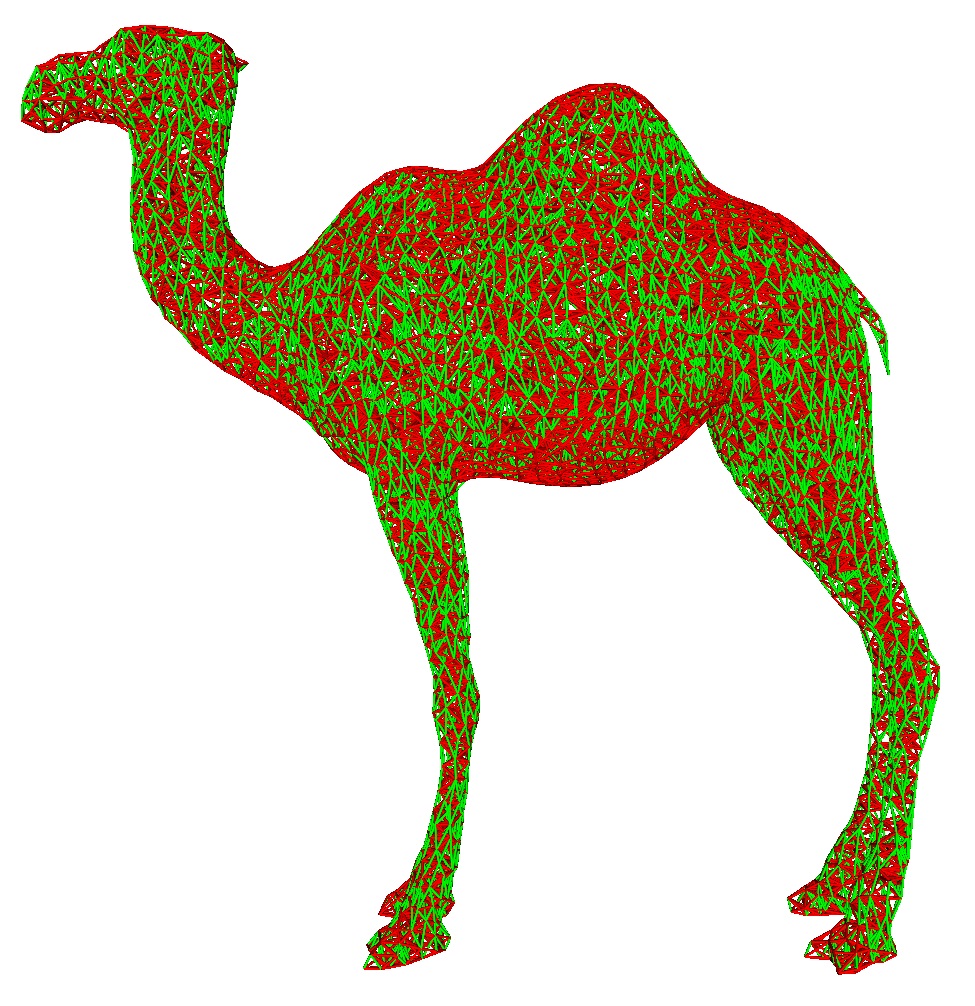}
\includegraphics[width=0.33\linewidth]{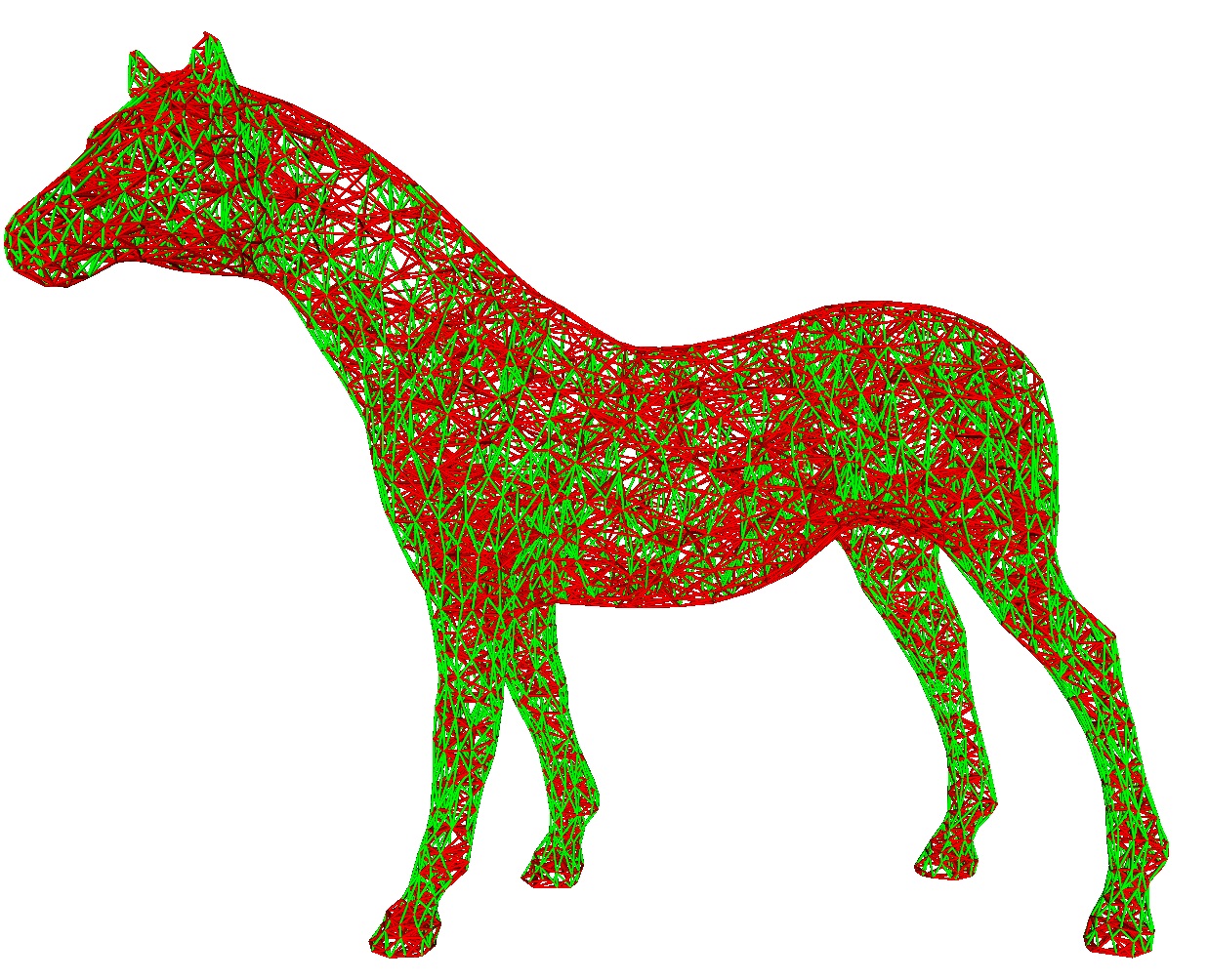}
}
\subfigure[Lattice structures generated by applying the SO and PO steps for self-supporting optimization in our approach -- the values of $\Psi$ are $47.3\%$ (Teddy), $48.6\%$ (Kitten), $45.5\%$ (Camel) and $49.7\%$ (Horse) respectively]{
\includegraphics[width=0.17\linewidth]{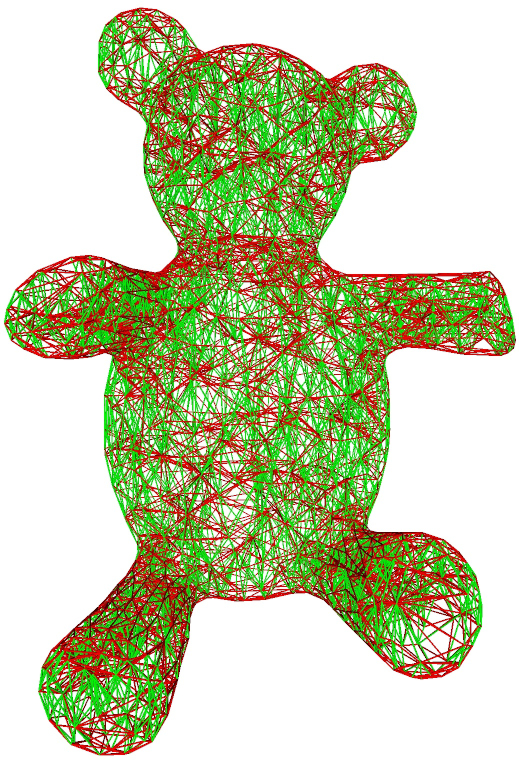}
\includegraphics[width=0.16\linewidth]{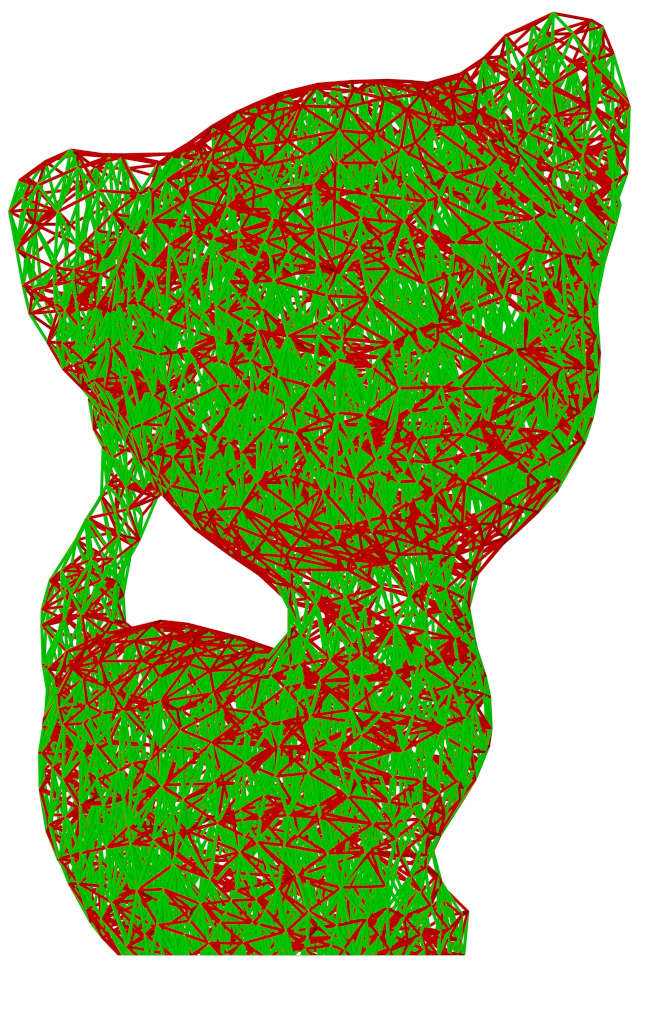}
\includegraphics[width=0.25\linewidth]{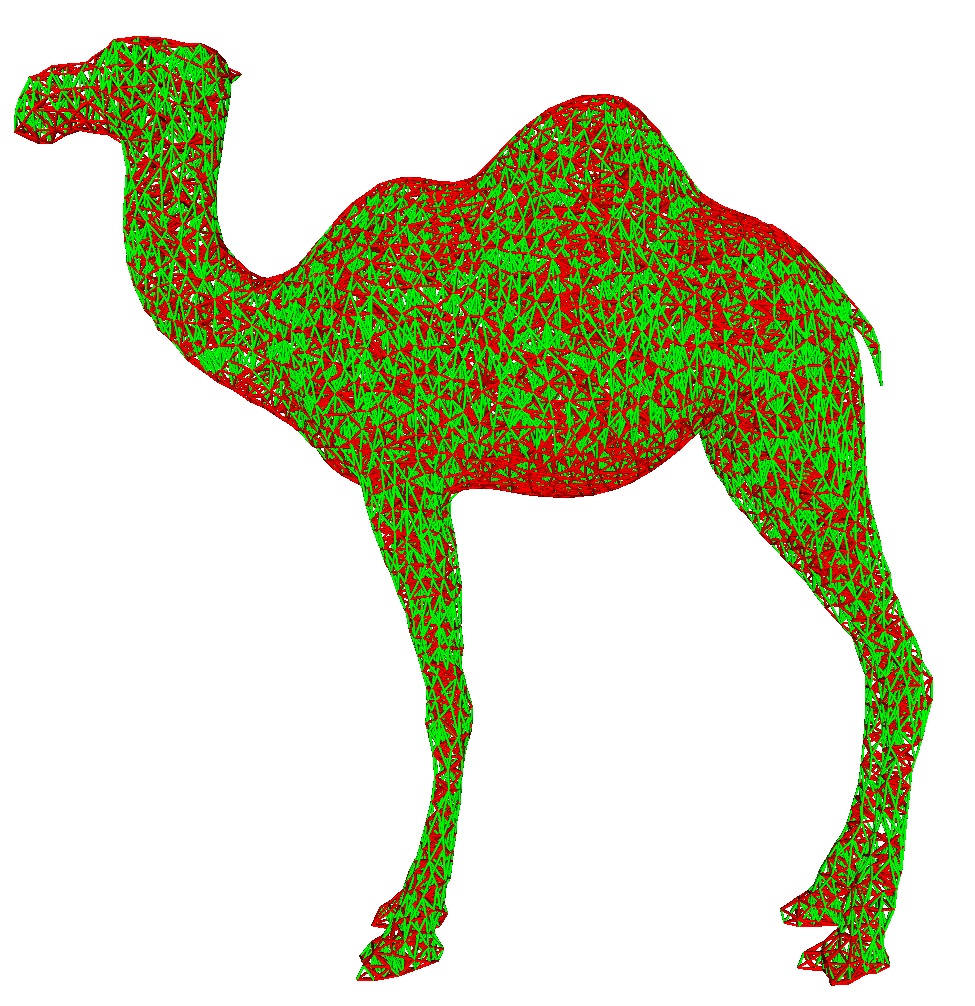}
\includegraphics[width=0.33\linewidth]{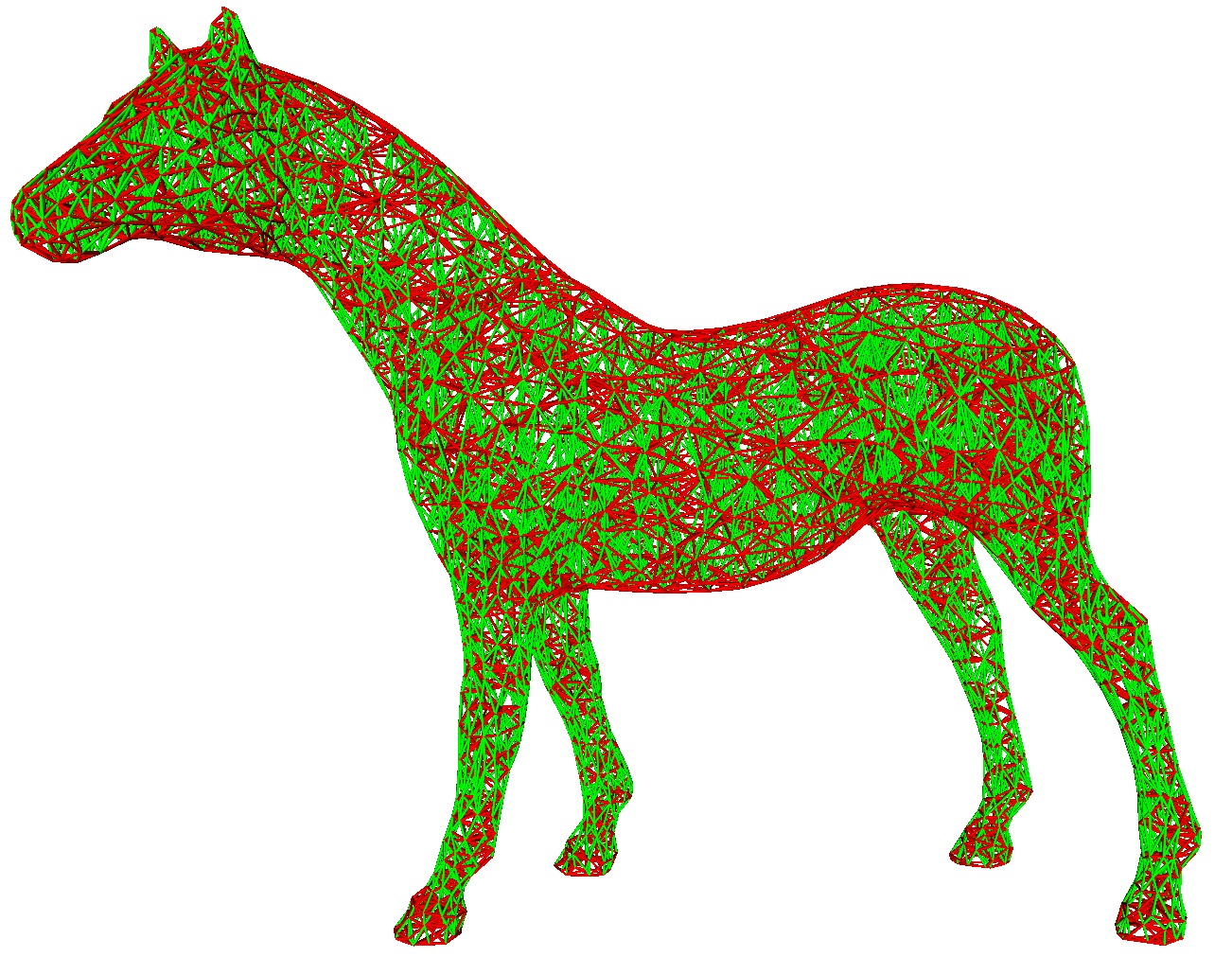}
}
\caption{Lattice structures as infills for four example models -- (from left to right) Teddy, Kitten, Camel and Horse, where the edges need additional supporting structures in 3D printing are displayed in red color. It is easy to find our results having much less number of `red' edges. The length percentage of completely self-supported struts $\Psi$ (Eq.(\ref{eqSupportLengthMetric})) are reported as well. The values of self-supporting metric $\Gamma$ (Eq.(\ref{eqSupportMetric}) are reported in Table \ref{tab:StatisticsOnSelfSupporting} for each model by using the constant radius for all struts. }
\label{fig:SelfSupporingRes}
\end{figure}

The comparison results are shown in Fig.\ref{fig:SelfSupporingRes}, where the first row gives the results of Tet-Gen, and the second row is our results. We also report the length percentage of self-supporting edges $\Psi$ (in Fig.\ref{fig:SelfSupporingRes}) and the values of self-supporting metric $\Gamma$ (in Table \ref{tab:StatisticsOnSelfSupporting}). Our method achieved a $50.8\%$ to $73.5\%$ improvement to the Tet-Gen method, as expected. In addition, from the computation time listed in the last column of Table \ref{tab:StatisticsOnSelfSupporting}, Our method is very fast, with all optimizations done within three minutes.

\begin{table}[t]
\centering 
\caption{The statistic for self-supporting optimization by measuring the values of $\Gamma(\Omega)$ in Eq.(\ref{eqSupportMetric})}\vspace{8pt}\label{tab:StatisticsOnSelfSupporting} 
\small
\begin{tabular}{l|c|c|c|c|c}
  \hline
  & & & \multicolumn{3}{c}{Our self-supporting optm.} \\
  \cline{4-6}
  \textbf{Model} & Fig. & Tet-Gen \cite{si2015tetgen} & SO& SO+PO & Time (sec.) \\
  \hline
  \hline
  Teddy & \ref{fig:SelfSupporingRes} & 59.7 & 48.8 & 30.1 & 171.3 \\
  \hline
  Kitten & \ref{fig:SelfSupporingRes} & 63.5 & 47.9 & 37.0 & 120.2  \\
  \hline
  Camel  & \ref{fig:SelfSupporingRes} & 63.1 & 50.0 & 37.3 & 89.3 \\
  \hline
  Horse  & \ref{fig:SelfSupporingRes} & 60.7 & 54.8 & 33.8 & 92.5 \\
  \hline
  Cube  & \ref{fig:OverviewSpatialGradedDensity} & 60.4 & 47.3 & 40.3 & 22.6 \\
  \hline
\end{tabular}
\begin{flushleft}
$^{*}$All models are evaluated by using the same radius for all struts. 
\end{flushleft}
\end{table}

\subsection{Slicing and Fabrication}\label{subsecResultFab}
\begin{figure}[t]
\includegraphics[width=\linewidth]{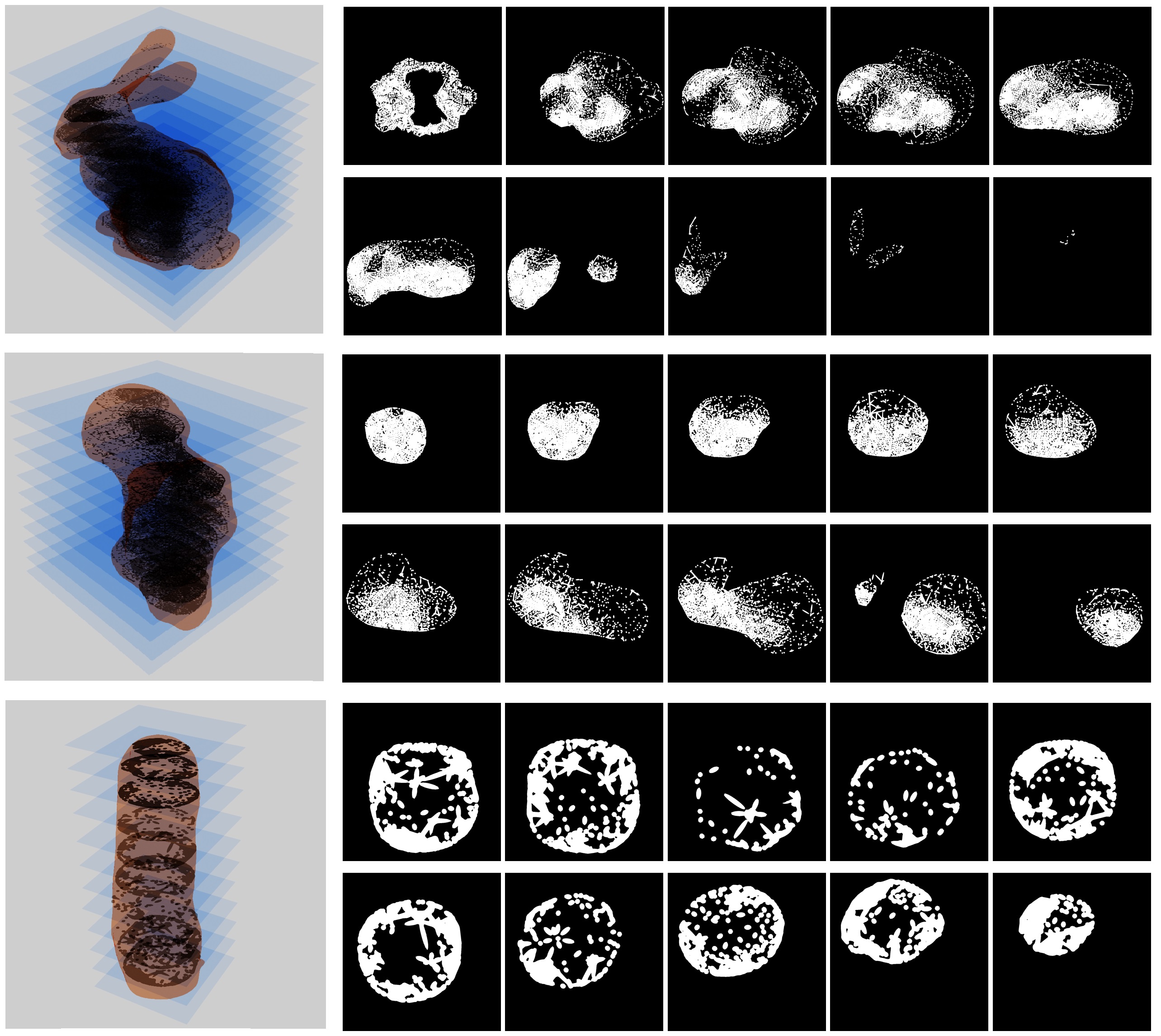}
\caption{The binary images obtained by slicing lattice structures (the ones shown in Fig.\ref{fig:DensityMatchingRes}) represented by our method.}
\label{fig:sliceimages}
\end{figure}
\begin{figure}[t]
\centering
\includegraphics[width=\linewidth]{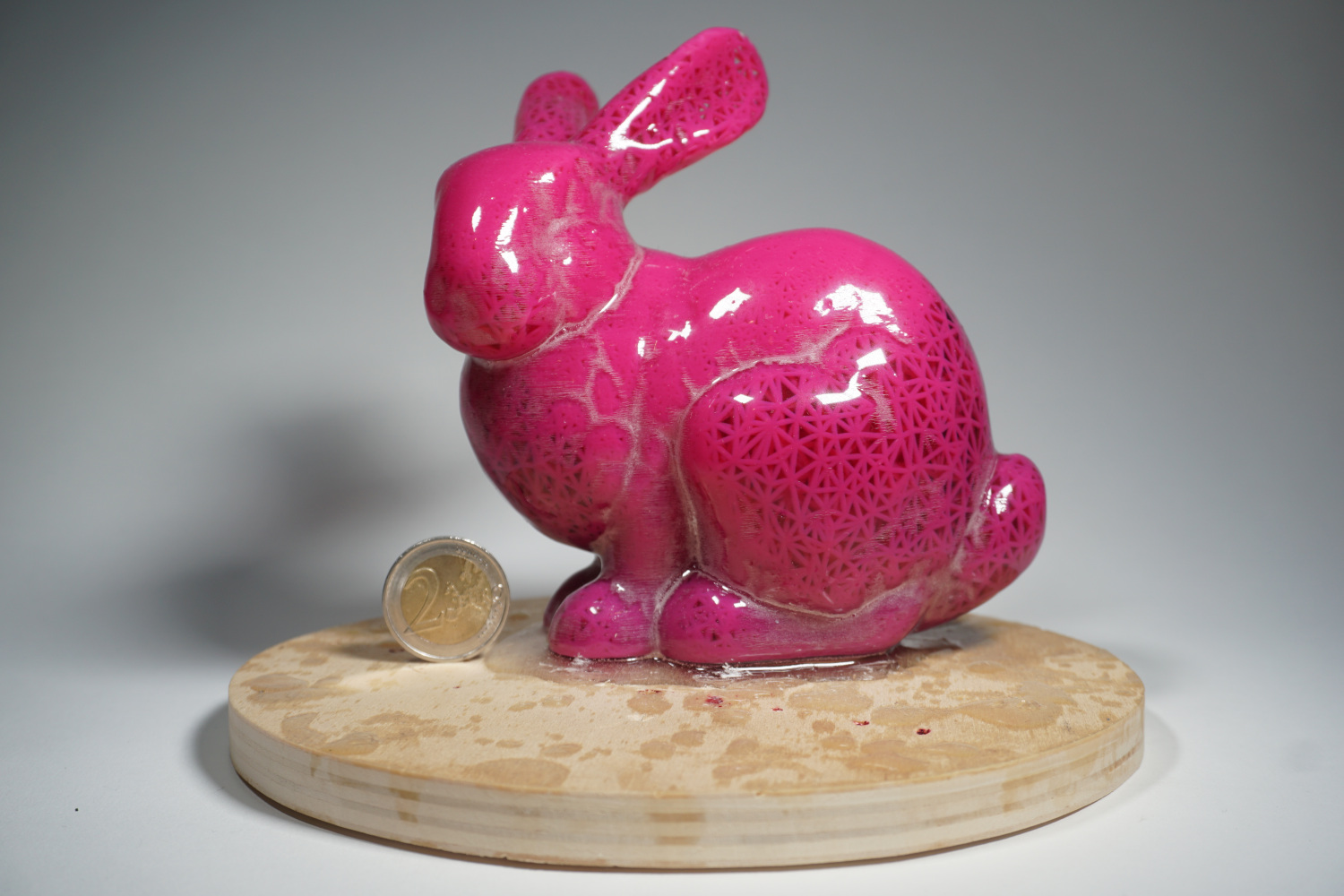}	
\caption{A bunny model with lattice structure generated by our approach as shown in Fig.\ref{fig:DensityMatchingRes} -- the model is fabricated by using a Connex Object350 3D printer.}
\label{fig:3DPrintedBunny}
\end{figure}

In this sub-section, we demonstrate the efficiency feature of our slicing method (Section~\ref{secSlicing}). All the models previously tested have been sliced using this algorithm. Fig.~\ref{fig:sliceimages} shows the resultant binary images of three of them: the bunny model, the bone model, and the finger model. Based on the binary images, we printed out the Bunny model to further validate our slicing algorithm, as shown in Fig.\ref{fig:3DPrintedBunny}. A DLP 3D printer was used, and the method presented in \cite{huang2014image} was chosen to generate supports wherever necessary.

\begin{figure}[t]
\centering
\includegraphics[width=\linewidth]{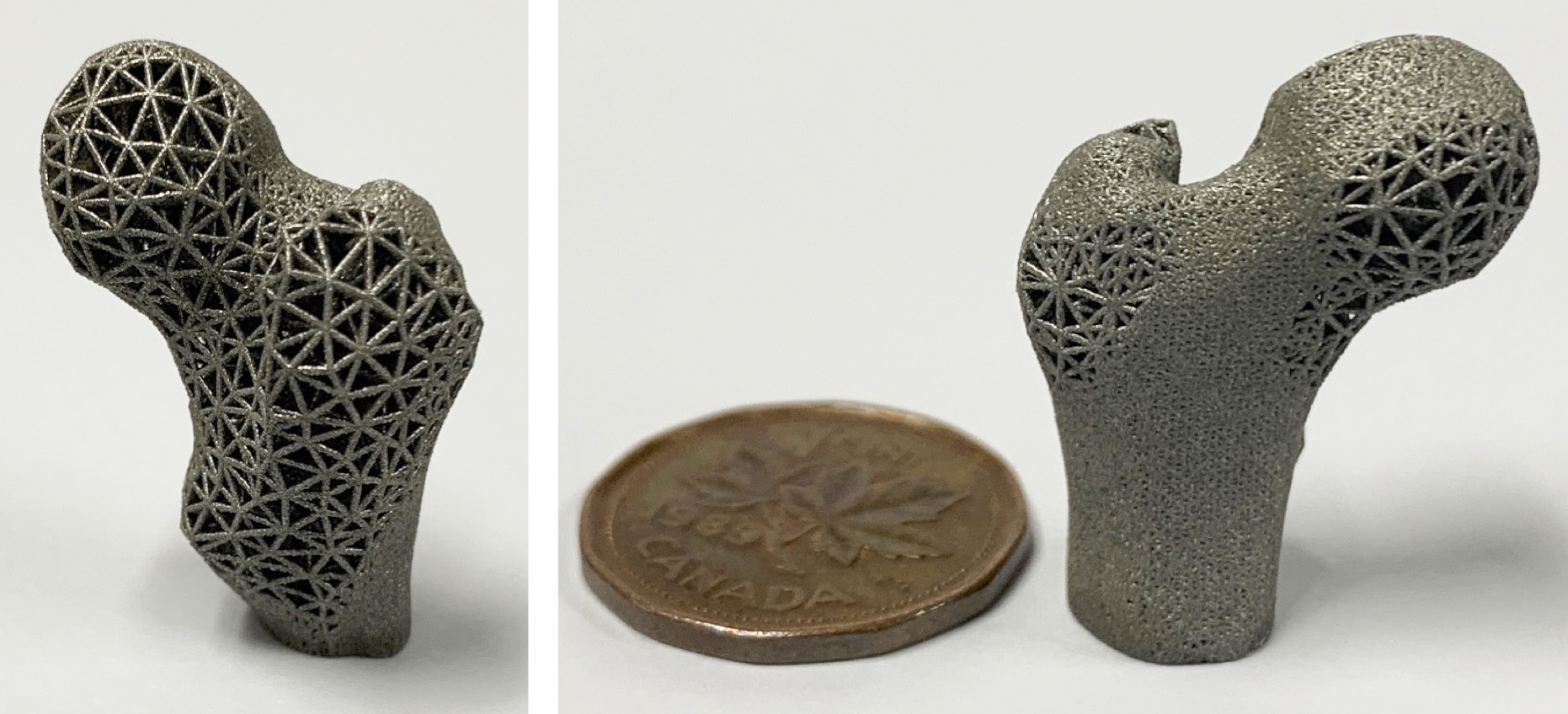}	
\caption{A bone model with lattice structure generated by our approach as shown in Fig.\ref{fig:DensityMatchingRes} -- the metal model is fabricated by a SLM 3D printer).}
\label{fig:3DPrintedBone}
\end{figure}

A result of bone model is fabricated by a \textit{Selective Laser Melting} (SLM) based metal 3D printer (as shown in Fig.\ref{fig:3DPrintedBone}) -- the dimensions are $19.2 \mathrm{mm} \times 12.6\mathrm{mm} \times 26.7\mathrm{mm}$. After generating the binary image for each slice, the contour of boundary is generated by the method of \cite{huang2013slicing} and the zigzag toolpath is employed to fill the interior region. The model was fabricated in $9.5$ hours. The printer has a 500W IPG fiber laser and a 25$\mu$m beam size.

Table \ref{tab:StatisticsOnSlicing} gives the memory consumption and the time usage statistics in slicing those models. As our method works in a streaming manner, the consumed memory and time is layer dependent, rather than model dependent. Specifically, the memory and time have a positive correlation with the maximal number of struts intersecting with a specific slicing plane, as confirmed by the statistics in Table \ref{tab:StatisticsOnSlicing}. For all the tested models, the slicing algorithm is observed to generate correct binary images fast and memory-efficiently.

\begin{figure}[t]
\centering
\includegraphics[width=\linewidth]{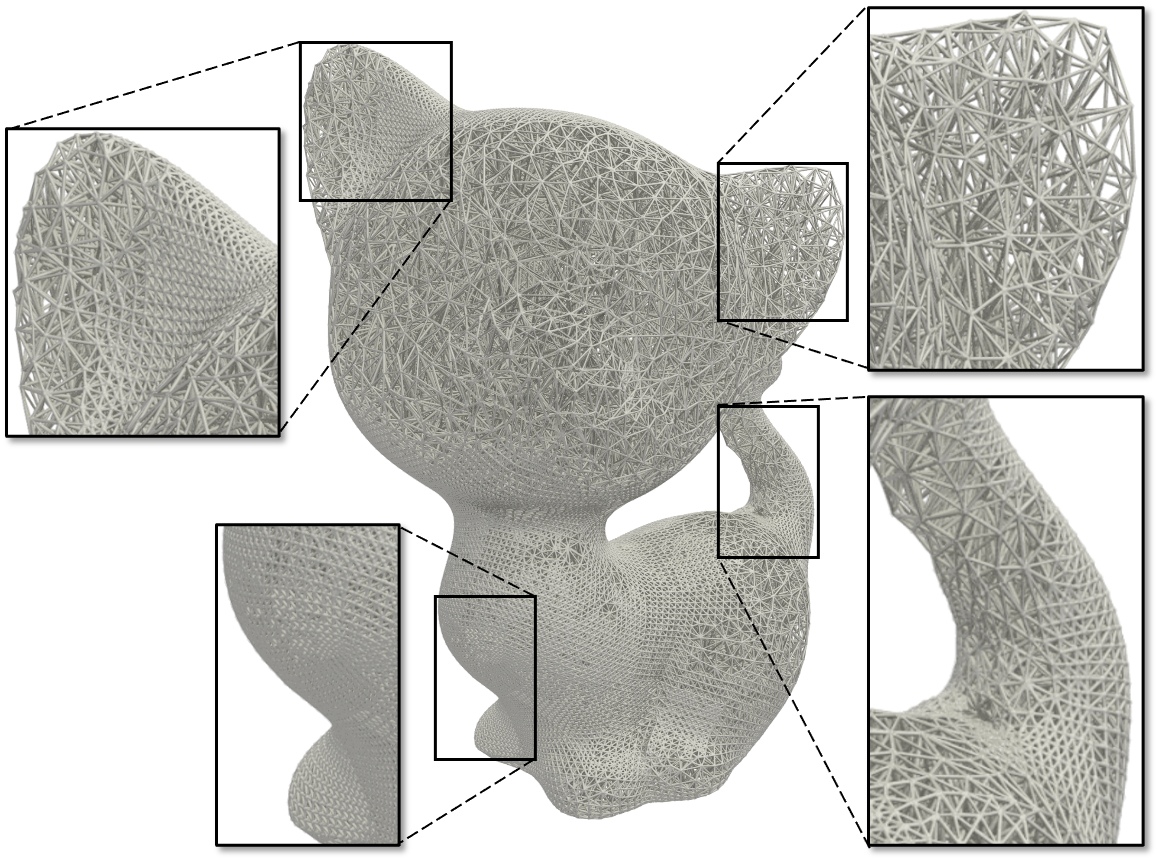}	
\caption{We are able to model and slice a lattice structure with millions of struts (14,063,027) by the streaming mode. Here the solid implicit model generated by our method is rendered by direct ray-tracing~\cite{plachetka1998pov}.}
\label{fig:KittenHR}
\end{figure}

Our approach is very scalable for models even with a huge number of struts. Although the size of models that can be 3D printed is currently limited by the hardware made available, we demonstrate the method's scalability by using a lattice structure with more than 101M struts (i.e., the Kitten-HR model as shown in Fig.\ref{fig:KittenHR}). When generating binary images for 3D printing, the maximal number of intersected struts is about 1M with the maximal memory-usage at 447MB. This fits quite well in a commercially available computer system.

\begin{table}[t]
\centering 
\caption{Statistic for our slicing algorithm}\label{tab:StatisticsOnSlicing}\vspace{8pt}
\small
\begin{tabular}{r|r|r|r|r}
  \hline
   & Max \# of & Used & Resolution & Time   \\
 \textbf{Model} & Intersected & Mem. &  of & (Sec.) \\
  (\# of Struts) & Struts & (MB) & Images & / Slice  \\
  \hline 
  \hline 
  Teddy & 1,720  & 22 & $1536 \times768$  & 0.10 \\
  (17,462) &       &     &           &      \\
  \hline
  Kitten & 38,439  & 88 & $1489\times1368$  & 18.82 \\
  (122,925) &       &     &           &      \\
  \hline
  Camel &  2,642 & 43 & $1456\times728$  & 1.10 \\
  (12,241) &          &     &           &      \\
  \hline
  Horse & 1,152 & 35 & $1592\times796$ & 0.09 \\
  (16,301) &         &     &           &      \\
  \hline
  Finger & 21,490 & 97 & $1808\times904$ & 15.63 \\
  (359,212) &        &     &           &      \\
  \hline
  Bone &  17,303 & 87 & $2768\times1384$ & 14.50  \\
  (75,484) &        &     &           &      \\
  \hline
  Bunny &  26,069 & 146 & $3648\times1824$ & 26.47 \\
  (123,610) &         &     &           &      \\
  \hline
  Kitten-HR & 1,061,866 & 447 & $5728\times2864$ & 347.23 \\
  (101,514,060) &     &     &           &      \\
  \hline
\end{tabular}
\end{table}

\section{Conclusions}
An implicit modeling technique is presented in this paper for large-scale adaptive lattice structures, which have a lot of applications in additive manufacturing. Starting from the edges of a graph, the solid of an initial lattice structure is defined using convolution surfaces with edges of the graph as skeletons. Different from the methods based on distance-field, solids defined by convolution surfaces are highly smooth at the knots with complex topology. This gives better mechanical strength than the solids with creases. Benefit from the local support formulation of convolution surface in our approach, the representation is very memory-efficient as only skeletons needs to be stored and the slicing of solids can be efficiently computed as only limited number of skeletons are involved in computing the intersection. This results in a highly scalable approach -- lattice structures with more than tens of million struts can be effectively modeled by our method.


The functionality of our approach has been demonstrated in the application to generate an adaptive lattice structure matching the given density distribution. The matched density is achieved by two operations: structural subdivision and strut-radius adaptation. The results are quite encouraging, where the desired densities are realized at all places inside a given 3D model. For those regions need to add supporting structures, the finally realized density on a physically fabricated model could be larger than the desired one when using single-material 3D printing. Although this will not reduce the mechanical strength in the model, we plan to model supporting structures by convolutional surface in an unique representation. As a consequence, the lattice structure fabricated by single-martial 3D printing can also match the desired density precisely.

Moreover, in the future work, we plan to use this method for designing lattice structures with different spatially graded physical properties such as a heat exchanger with large surface area within a small volume, an energy absorber tolerating great deformation at a low stress level and an acoustic insulator with its large number of internal pores. In all these application, the convolution surface based modelling method proposed in our work can show great advantages in its effectiveness and scalability.

\begin{acknowledgment}
This research work is partially supported by the Natural Science Foundation of China (NSFC) (Project Ref. No.: 61628211, 61572527) and the Hunan Science Fund for Distinguished Young Scholars (Ref. No.: 2019JJ20027). The authors would like to acknowledge the private communication with Jun Wu during the development of this project.
\end{acknowledgment}

%
% ...

\bibliographystyle{asmems4}

\bibliography{asme2e}

\begin{thebibliography}{10}

\bibitem{martinez2016procedural}
Mart{\'\i}nez, J., Dumas, J., and Lefebvre, S., 2016.
\newblock ``Procedural voronoi foams for additive manufacturing''.
\newblock {\em ACM Transactions on Graphics (TOG), {\bf 35}}(4), p.~44.

\bibitem{martinez2017orthotropic}
Mart{\'\i}nez, J., Song, H., Dumas, J., and Lefebvre, S., 2017.
\newblock ``Orthotropic k-nearest foams for additive manufacturing''.
\newblock {\em ACM Transactions on Graphics (TOG), {\bf 36}}(4), p.~121.

\bibitem{kuipers2019CrossFill}
Kuipers, T., Wu, J., and Wang, C.~C., 2019.
\newblock ``{CrossFill}: Foam structure with graded density for continuous
  material extrusion''.
\newblock {\em Computer-Aided Design, {\bf 114}}, pp.~37--50.

\bibitem{qin2017mechanics}
Qin, Z., Jung, G.~S., Kang, M.~J., and Buehler, M.~J., 2017.
\newblock ``The mechanics and design of a lightweight three-dimensional
  graphene assembly''.
\newblock {\em Science Advances, {\bf 3}}(1), p.~e1601536.

\bibitem{rosen2006lattice}
Rosen, D., Johnston, S., Reed, M., and Wang, H.
\newblock ``Design of general lattice structures for lightweight and compliance
  applications''.
\newblock In Rapid Manufacturing Conference, July 5-6, 2006, Loughborough
  University.

\bibitem{chen2008LDNI}
Chen, Y., and Wang, C.~C.
\newblock ``Layered depth-normal images for complex geometries - part one:
  accurate sampling and adaptive modeling''.
\newblock In ASME IDETC/CIE 2008 Conference, 28th Computers and Information in
  Engineering Conference, New York City, New York, August 3-6, 2008.

\bibitem{chougrani2017lattice}
Chougrani, L., Pernot, J.-P., V{\'e}ron, P., and Abed, S., 2017.
\newblock ``Lattice structure lightweight triangulation for additive
  manufacturing''.
\newblock {\em Computer-Aided Design, {\bf 90}}, pp.~95--104.

\bibitem{si2015tetgen}
Si, H., 2015.
\newblock ``Tetgen, a delaunay-based quality tetrahedral mesh generator''.
\newblock {\em ACM Transactions on Mathematical Software (TOMS), {\bf 41}}(2),
  p.~11.

\bibitem{gao2015status}
Gao, W., Zhang, Y., Ramanujan, D., Ramani, K., Chen, Y., Williams, C.~B., Wang,
  C.~C., Shin, Y.~C., Zhang, S., and Zavattieri, P.~D., 2015.
\newblock ``The status, challenges, and future of additive manufacturing in
  engineering''.
\newblock {\em Computer-Aided Design, {\bf 69}}, pp.~65--89.

\bibitem{livesu20173d}
Livesu, M., Ellero, S., Mart{\'\i}nez, J., Lefebvre, S., and Attene, M., 2017.
\newblock ``From {3D} models to {3D} prints: an overview of the processing
  pipeline''.
\newblock {\em Computer Graphics Forum, {\bf 36}}(2), pp.~537--564.

\bibitem{Leung2019}
Leung, Y.-S., Kwok, T.-H., Li, X., Yang, Y., Wang, C. C.~L., and Chen, Y.,
  2019.
\newblock ``Challenges and status on design and computation for emerging
  additive manufacturing technologies''.
\newblock {\em Journal of Computing and Information Science in Engineering,
  {\bf 19}}(2), 03.
\newblock 021013.

\bibitem{ding2014tool}
Ding, D., Pan, Z.~S., Cuiuri, D., and Li, H., 2014.
\newblock ``A tool-path generation strategy for wire and arc additive
  manufacturing''.
\newblock {\em International Journal of Advanced Manufacturing Technology, {\bf
  73}}(1-4), pp.~173--183.

\bibitem{zhao2016connected}
Zhao, H., Gu, F., Huang, Q.-X., Garcia, J., Chen, Y., Tu, C., Benes, B., Zhang,
  H., Cohen-Or, D., and Chen, B., 2016.
\newblock ``Connected fermat spirals for layered fabrication''.
\newblock {\em ACM Transactions on Graphics (TOG), {\bf 35}}(4), p.~100.

\bibitem{steuben2016implicit}
Steuben, J.~C., Iliopoulos, A.~P., and Michopoulos, J.~G., 2016.
\newblock ``Implicit slicing for functionally tailored additive
  manufacturing''.
\newblock {\em Computer-Aided Design, {\bf 77}}, pp.~107--119.

\bibitem{kumar2009fractal}
Kumar, G.~S., Pandithevan, P., and Ambatti, A.~R., 2009.
\newblock ``Fractal raster tool paths for layered manufacturing of porous
  objects''.
\newblock {\em Virtual and Physical Prototyping, {\bf 4}}(2), pp.~91--104.

\bibitem{wu2016self}
Wu, J., Wang, C.~C., Zhang, X., and Westermann, R., 2016.
\newblock ``Self-supporting rhombic infill structures for additive
  manufacturing''.
\newblock {\em Computer-Aided Design, {\bf 80}}, pp.~32--42.

\bibitem{lee2017block}
Lee, J., and Lee, K., 2017.
\newblock ``Block-based inner support structure generation algorithm for 3d
  printing using fused deposition modeling''.
\newblock {\em The International Journal of Advanced Manufacturing Technology,
  {\bf 89}}(5-8), pp.~2151--2163.

\bibitem{lu2014build}
Lu, L., Sharf, A., Zhao, H., Wei, Y., Fan, Q., Chen, X., Savoye, Y., Tu, C.,
  Cohen-Or, D., and Chen, B., 2014.
\newblock ``Build-to-last: strength to weight 3d printed objects''.
\newblock {\em ACM Transactions on Graphics (TOG), {\bf 33}}(4), p.~97.

\bibitem{lee2018support}
Lee, M., Fang, Q., Cho, Y., Ryu, J., Liu, L., and Kim, D.-S., 2018.
\newblock ``Support-free hollowing for 3d printing via voronoi diagram of
  ellipses''.
\newblock {\em Computer-Aided Design, {\bf 101}}, pp.~23--36.

\bibitem{Stankovic2020}
Stankovi\'{c}, T., and Shea, K., 2020.
\newblock ``Investigation of a {V}oronoi diagram representation for the
  computational design of additively manufactured discrete lattice
  structures''.
\newblock {\em Journal of Mechanical Design, {\bf 142}}(11), 05.
\newblock 111704.

\bibitem{wang2013cost}
Wang, W., Wang, T.~Y., Yang, Z., Liu, L., Tong, X., Tong, W., Deng, J., Chen,
  F., and Liu, X., 2013.
\newblock ``Cost-effective printing of 3d objects with skin-frame structures''.
\newblock {\em ACM Transactions on Graphics (TOG), {\bf 32}}(6), p.~177.

\bibitem{zhang2015medial}
Zhang, X., Xia, Y., Wang, J., Yang, Z., Tu, C., and Wang, W., 2015.
\newblock ``Medial axis tree—an internal supporting structure for 3d
  printing''.
\newblock {\em Computer Aided Geometric Design, {\bf 35}}, pp.~149--162.

\bibitem{schumacher2015microstructures}
Schumacher, C., Bickel, B., Rys, J., Marschner, S., Daraio, C., and Gross, M.,
  2015.
\newblock ``Microstructures to control elasticity in 3d printing''.
\newblock {\em ACM Transactions on Graphics (TOG), {\bf 34}}(4), p.~136.

\bibitem{panetta2015elastic}
Panetta, J., Zhou, Q., Malomo, L., Pietroni, N., Cignoni, P., and Zorin, D.,
  2015.
\newblock ``Elastic textures for additive fabrication''.
\newblock {\em ACM Transactions on Graphics (TOG), {\bf 34}}(4), p.~135.

\bibitem{fryazinov2013multi}
Fryazinov, O., Vilbrandt, T., and Pasko, A., 2013.
\newblock ``Multi-scale space-variant frep cellular structures''.
\newblock {\em Computer-Aided Design, {\bf 45}}(1), pp.~26--34.

\bibitem{yang2018computing}
Yang, Y., Chai, S., and Fu, X.-M., 2018.
\newblock ``Computing interior support-free structure via hollow-to-fill
  construction''.
\newblock {\em Computers \& Graphics, {\bf 70}}, pp.~148--156.

\bibitem{christiansen2015automatic}
Christiansen, A.~N., Schmidt, R., and B{\ae}rentzen, J.~A., 2015.
\newblock ``Automatic balancing of {3D} models''.
\newblock {\em Computer-Aided Design, {\bf 58}}, pp.~236--241.

\bibitem{stava2012stress}
Stava, O., Vanek, J., Benes, B., Carr, N., M{\`e}s, R., J{\'e}r{\'e}mie, and
  Lefebvre, S., 2012.
\newblock ``Stress relief: improving structural strength of 3d printable
  objects''.
\newblock {\em ACM Transactions on Graphics (TOG), {\bf 31}}(4), pp.~48:1--11.

\bibitem{ou2016cilllia}
Ou, J., Dublon, G., Cheng, C.-Y., Heibeck, F., Willis, K., and Ishii, H., 2016.
\newblock ``Cilllia: 3d printed micro-pillar structures for surface texture,
  actuation and sensing''.
\newblock In Proceedings of the 2016 CHI Conference on Human Factors in
  Computing Systems, pp.~5753--–5764.

\bibitem{vanek2014clever}
Vanek, J., Galicia, J. A.~G., and Benes, B., 2014.
\newblock ``Clever support: Efficient support structure generation for digital
  fabrication''.
\newblock In Computer graphics forum, Vol.~33, Wiley Online Library,
  pp.~117--125.

\bibitem{zhang2015perceptual}
Zhang, X., Le, X., Panotopoulou, A., Whiting, E., and Wang, C.~C., 2015.
\newblock ``Perceptual models of preference in 3d printing direction''.
\newblock {\em ACM Transactions on Graphics (TOG), {\bf 34}}(6), p.~215.

\bibitem{dumas2014bridging}
Dumas, J., Hergel, J., and Lefebvre, S., 2014.
\newblock ``Bridging the gap: automated steady scaffoldings for 3d printing''.
\newblock {\em ACM Transactions on Graphics (TOG), {\bf 33}}(4), p.~98.

\bibitem{hu2015support}
Hu, K., Jin, S., and Wang, C.~C., 2015.
\newblock ``Support slimming for single material based additive
  manufacturing''.
\newblock {\em Computer-Aided Design, {\bf 65}}, pp.~1--10.

\bibitem{wang2003nonmanifold}
Wang, C.~C., Wang, Y., and Yuen, M.~M., 2003.
\newblock ``Feature-based 3d non-manifold freeform object construction''.
\newblock {\em Engineering with Computers, {\bf 19}}, pp.~174--190.

\bibitem{sherstyuk1999kernel}
Sherstyuk, A., 1999.
\newblock ``Kernel functions in convolution surfaces: a comparative analysis''.
\newblock {\em The Visual Computer, {\bf 15}}, pp.~171--–182.

\bibitem{hubert2012convolution}
Hubert, E., and Cani, M.-P., 2012.
\newblock ``Convolution surfaces based on polygonal curve skeletons''.
\newblock {\em Journal of Symbolic Computation, {\bf 47}}(6), pp.~680–--699.

\bibitem{Tang2019}
Tang, Y., Xiong, Y., in~Park, S., Boddeti, G.~N., and Rosen, D.~W., 2019.
\newblock ``Generation of lattice structures with convolution surface''.
\newblock In Proceedings of CAD'19 Conference, pp.~69--74.

\bibitem{jin2002convolution}
Jin, X., and Tai, C.-L., 2002.
\newblock ``Analytical methods for polynomial weighted convolution surfaces
  with various kernels''.
\newblock {\em Computers \& Graphics, {\bf 26}}(3), pp.~437--447.

\bibitem{Nelaturi2015}
Nelaturi, S., and Shapiro, V., 2015.
\newblock ``Representation and analysis of additively manufactured parts''.
\newblock {\em Computer-Aided Design, {\bf 67}}, p.~13–23.

\bibitem{larsen1999fast}
Larsen, E., Gottschalk, S., Lin, M.~C., and Manocha, D., 1999.
\newblock Fast proximity queries with swept sphere volumes.
\newblock Tech. rep., Technical Report TR99-018, Department of Computer
  Science, University of North Carolina at Chapel Hill.

\bibitem{hearn2004computer}
Hearn, D., Baker, M.~P., et~al., 2004.
\newblock {\em Computer graphics with OpenGL}.
\newblock Upper Saddle River, NJ: Pearson Prentice Hall,.

\bibitem{Mcmains2000}
Mc{M}ains, S.~A., 2000.
\newblock ``Geometric algorithms and data representation for solid freeform
  fabrication''.
\newblock PhD thesis.

\bibitem{huang2014image}
Huang, P., Wang, C.~C., and Chen, Y., 2014.
\newblock ``Algorithms for layered manufacturing in image space''.
\newblock In {ASME} Advances in Computers and Information in Engineering
  Research, Vol.~1, pp.~377--410.

\bibitem{maple2003ms}
{Maple}, C., 2003.
\newblock ``Geometric design and space planning using the marching squares and
  marching cube algorithms''.
\newblock In 2003 International Conference on Geometric Modeling and Graphics,
  2003. Proceedings, pp.~90--95.

\bibitem{huang2013slicing}
Huang, P., Wang, C. C.~L., and Chen, Y., 2013.
\newblock ``Intersection-free and topologically faithful slicing of implicit
  solid''.
\newblock {\em Journal of Computing and Information Science in Engineering,
  {\bf 13}}(2), 04.
\newblock 021009.

\bibitem{surazhsky2003remeshing}
Surazhsky, V., and Gotsman, C., 2003.
\newblock ``Explicit surface remeshing''.
\newblock In Proceedings of the 2003 Eurographics/ACM SIGGRAPH Symposium on
  Geometry Processing, Eurographics Association, pp.~20--30.

\bibitem{lorensen1987marching}
Lorensen, W.~E., and Cline, H.~E., 1987.
\newblock ``Marching cubes: A high resolution 3d surface construction
  algorithm''.
\newblock In ACM siggraph computer graphics, Vol.~21, ACM, pp.~163--169.

\bibitem{plachetka1998pov}
Plachetka, T., 1998.
\newblock ``Pov ray: persistence of vision parallel raytracer''.
\newblock In Proc. of Spring Conf. on Computer Graphics, Budmerice, Slovakia,
  Vol.~123.

\end{thebibliography}

\appendix       


\section*{Appendix I: Estimate Target Edge-Length of Tetrahedron}
\begin{figure}[t]
\centering
\includegraphics[width=\linewidth]{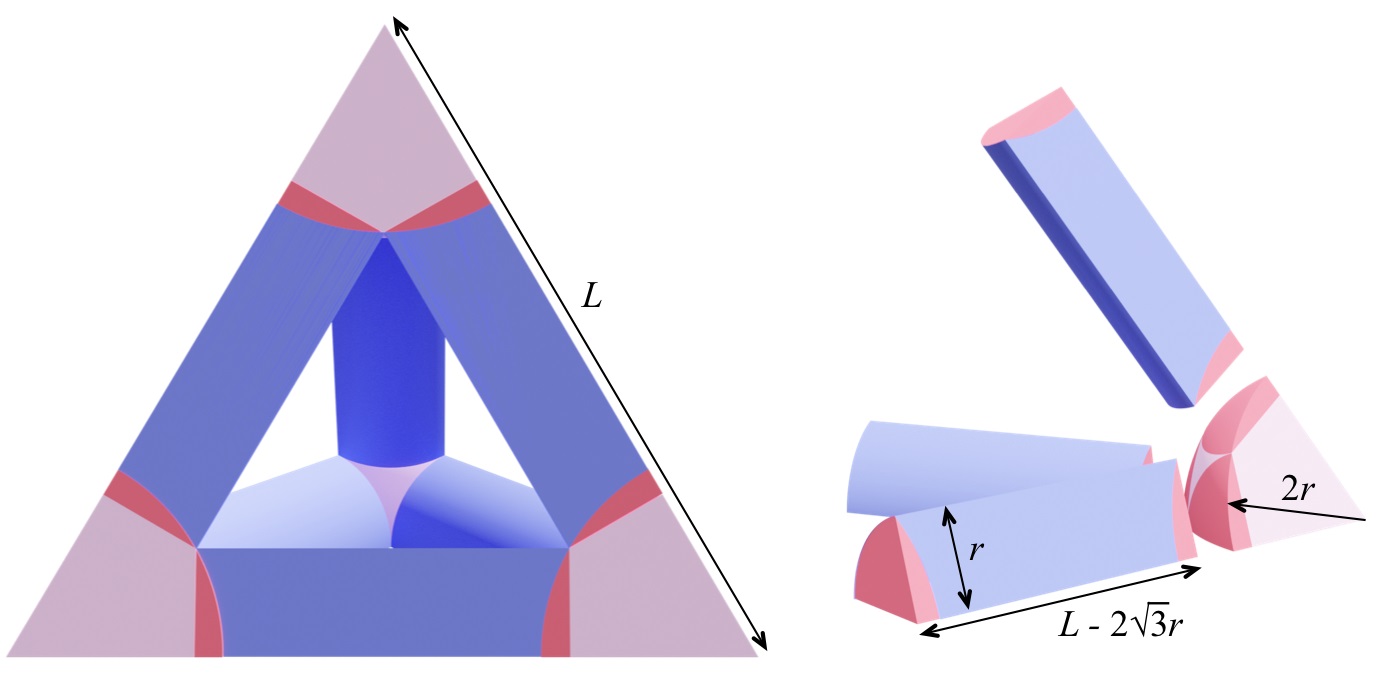}	
\caption{The volume of the lattice structure inside a regular tetrahedron can be approximately evaluated by the decomposition of 6 cylinders (with length $L$ and radius $r$) and 4 spheres (with radius $2r$), where the red regions are overlapped so that leads to approximation errors.}
\label{fig:Appendix}
\end{figure}

Given the edge length of a regular tetrahedron as $L$, the tetrahedron's volume is
\begin{equation}
   V_{tet} = \frac{\sqrt{2}}{12}L^3
\end{equation}
We then calculate the volume of beams inside the tetrahedron, which consists of two parts -- the cylindrical regions and the spherical regions (see Fig.\ref{fig:Appendix} for an illustration).

\begin{enumerate}
\item The cylinder volume of an edge with length $L$ and radius $r$ is $L r^2 \pi$, which has the volume $\sqrt{3}r^3 \pi$ overlapped with two spheres with radius $r$ centered at the edge's two endpoints. The dihedral angle of every 2 faces of tetrahedron is $\operatorname{Arccos}\left(\frac{1}{3}\right)$. A tetrahedron has six edges, thus we have the volume of the cylindrical part inside the tetrahedron
\begin{center}
     $V_{cylinders} = 6\operatorname{Arccos} \left(\frac{1}{3}\right)(L-2\sqrt{3} r) r^{2}$
\end{center}

\item The sphere at a vertex has the approximate volume $V_{sphere} = 4 \sqrt{3} \pi r^3$. A tetrahedron has 4 vertices, and the steradian $\Omega$ of each vertex could be calculated as  
\begin{center}
$\Omega=\alpha+\beta+\gamma-\pi=3 \operatorname{Arccos}\left(\frac{1}{3}\right)-\pi$,
\end{center}
where $\alpha$, $\beta$ and $\gamma$ are dihedral angles of a vertex.
We can have the total volume at all the corners as
\begin{center}
     $V_{corners}=\frac{32}{3}\left(3\operatorname{Arccos}\left(\frac{1}{3}\right)-\pi\right) r^{3}$.
\end{center}
\end{enumerate}
When using material with density $\tau$ to realize the target density $\rho$ inside the tetrahedron, we should have
\begin{center}
    $\rho V_{tet} = \tau (V_{cylinders} + V_{corners})$.    
\end{center}
This leads to a density estimating formula as 
\begin{center}
    $\rho = 2\sqrt{2}\left(\frac{9\operatorname{Arccos}\left(\frac{1}{3}\right)r^2}{L^2}-\frac{\left(\left(96-18\sqrt{3}\right)\operatorname{Arccos}\left(\frac{1}{3}\right)-32\pi\right) r^3}{L^3}\right)\tau.$
\end{center}
Note that, the volume of merged struts estimated in the above way has some errors as the volume in overlapped regions of sphere and cylinders are double counted (see the red region shown in Fig.\ref{fig:Appendix}). However, as the purpose of this estimation is only to generate a target length for remeshing, this approximation will not influence the final result of density matching. The volume of $\mathcal{S}(\Omega)$ in our density matching framework is computed by Monte-Carlo integral with reference to the implicit solid.

When the target density $\rho$ is given, the formula can be rewritten into a cubic polynomial equation to estimate the target edge-length of the surface mesh,
\begin{equation}\label{eqMeshEdgeLengthEstimation}
    L^{3}+p L+q=0,
\end{equation}
where 
\begin{center}
    $p=-18 \sqrt{2}\operatorname{Arccos}\left(\frac{1}{3}\right) \tau r^{2} / \rho$, 
\end{center}
and 
\begin{center}
$q =-\left(\left(192\sqrt{2}-36\sqrt{6}\right)\operatorname{Arccos}\left(\frac{1}{3}\right)-64\sqrt{2}\pi\right) \tau r^3 / \rho$. 
\end{center}
According to Cardano formula, and it is easy to approve the discriminant for the roots $\Delta = \left(\frac{q}{2}\right)^2+\left(\frac{p}{3}\right)^3 < 0$, so this equation has three different real solutions:
\begin{center}
    $L_1 = \left(h_1\right)^\frac{1}{3}+\left(h_2\right)^\frac{1}{3}$,\\
    $L_2=\omega\left(h_1\right)^\frac{1}{3}+\omega^2\left(h_2\right)^\frac{1}{3}$,\\
    $L_3=\omega^2\left(h_1\right)^\frac{1}{3}+\omega\left(h_2\right)^\frac{1}{3}$,
\end{center}
where $h_1=-\frac{q}{2}+\sqrt{\left(\frac{q}{2}\right)^2+\left(\frac{p}{3}\right)^3}$, $h_2=-\frac{q}{2}-\sqrt{\left(\frac{q}{2}\right)^2+\left(\frac{p}{3}\right)^3}$, $\omega = \frac{-1+\sqrt(3)i}{2}$. Assuming the average edge-length $L_{cur}$ of the current mesh, $L_{ini}$, the one in $\{L_1, L_2, L_3\}$ being closest to $L_{cur}$ will be selected as a possible estimation of the target edge-length $\Bar{L}$. And in order to avoiding the vanish of an edge, the edge-length should be not less than $4r$. In short, we can have the following solution for $\Bar{L}$  
\begin{center}
    $\Bar{L} = \max \left\{ 4r , L_{ini} \right\}$.
\end{center}

\section*{Appendix II: Ratio of Risky Projected Area}
In this appendix, we derive the formula to calculate the ratio of risky projected area on a cylinder that needs additional supporting structure for 3D printing. All the analysis is conducted on a cylinder with unit radius, unit length and bottom-circle's center located at the origin $\mathbf{o}$. In its initial configuration, the cylinder's axis is aligned with the $z$-axis. Then, the parametric representation for a point on the bottom-circle as $\mathbf{q}(\phi)=(\cos \phi, \sin \phi, 0)$. 

Without loss of the generality, any strut with the angle $\theta$ between its axis and the printing direction $\mathbf{t}_P$ (see the illustration in Fig.\ref{fig:scaling4SelfSupporting}) can be considered as rotating around $x$- and $z$-axes and scaling the unit cylinder. Only rotating around $x$-axis will change the ratio of projected area that needs to add supporting structures. Considering the rotation matrix around $x$-axis as $\mathbf{R}_x(\theta)$, a point on the bottom-circle becomes
\begin{center}
$\mathbf{p}(\phi) = \mathbf{R}_x(\theta) \mathbf{q}(\phi) = (\cos \phi, \sin \phi \cos \theta, \sin \phi \sin \theta)$.
\end{center}
Therefore, the surface normal of any point on this circle is $\mathbf{n}=\mathbf{p}(\phi)-\mathbf{o}=\mathbf{p}(\phi)$. 

Considering the condition to add support as 
\begin{center}
$\mathbf{n} \cdot \mathbf{t}_P < -\sin \alpha$, 
\end{center}
we can determine the portion on the circle to add support by determine the range of $\phi$ that make 
\begin{center}
$\mathbf{n} \cdot \mathbf{t}_P = \mathbf{p}(\phi)  \cdot \mathbf{t}_P = \sin \phi \sin \theta < -\sin \alpha$.
\end{center}
Here we apply $\mathbf{t}_P = (0,0,1)$.

Now we project the circle back onto the $xy$-plane. For any $\theta \in (0,\pi / 2)$, projection of the bottom-circle gives an ellipse with width $a=1$ and height $b=\cos \theta$. The critical point that changes from self-supporting to support-needed can then determined by solving the elliptic equation and the equation embedding $\mathbf{n} \cdot \mathbf{t}_P = -\sin \alpha$, which can be 
\begin{center}
$\mathbf{p}_x^2 + \mathbf{p}_y^2 +\sin^{2}\alpha=\cos ^{2} \phi+\cos ^{2} \theta \sin ^{2} \phi +\sin ^{2} \theta \sin ^{2} \phi =1.$
\end{center}
Therefore, we have
\begin{center}
$\left\{\begin{array}{l}\mathbf{p}_x^2 + \frac{\mathbf{p}_y^2}{\cos^{2}\theta} = 1, \\
\mathbf{p}_x^2 + \mathbf{p}_y^2 = 1 - \sin^{2}\alpha.\end{array}\right.$
\end{center} 
By elimination, we obtain 
\begin{center}
$\left( \cos^{2}\theta-1\right) \mathbf{p}_x^2 = \cos^{2}\theta - 1 + \sin^{2}\alpha,$ \\
$\left(1-\frac{1}{ \cos^{2}\theta}\right) \mathbf{p}_y^2 = - \frac{\sin^{2}\theta}{ \cos^{2}\theta} \mathbf{p}_y^2 = -\sin^{2}\alpha.$
\end{center}
As a result, the solution of $\phi$ as $\phi_0$ that satisfies the above two equations can be obtained when the values of $\theta$ and $\alpha$ are given. In short, we have $\mathbf{p}_y = \frac{\sin \alpha \cos \theta}{\sin \theta} = \sin \phi_0 \cos \theta $, which results in the value of $\phi_0$ as
\begin{center}
$\phi_0 = \arcsin{(\frac{\sin{\alpha}}{\sin{\theta}})}$.
\end{center}
Since $\mathbf{p}_x^2 \geq 0$ and $(\cos^{2}\theta-1) < 0$, we should let $\cos^{2}\theta - 1 + \sin^{2}\alpha < 0$ to ensure there is a solution for $\mathbf{p}_x$. This leads to $\sin^2 \alpha \leq 1 - \cos^2\theta = \sin^2 \theta$, which actually requires $\theta > \alpha$ for risky area (i.e., area needs additional support). 

\begin{figure}
\centering
\includegraphics[width=.8\linewidth]{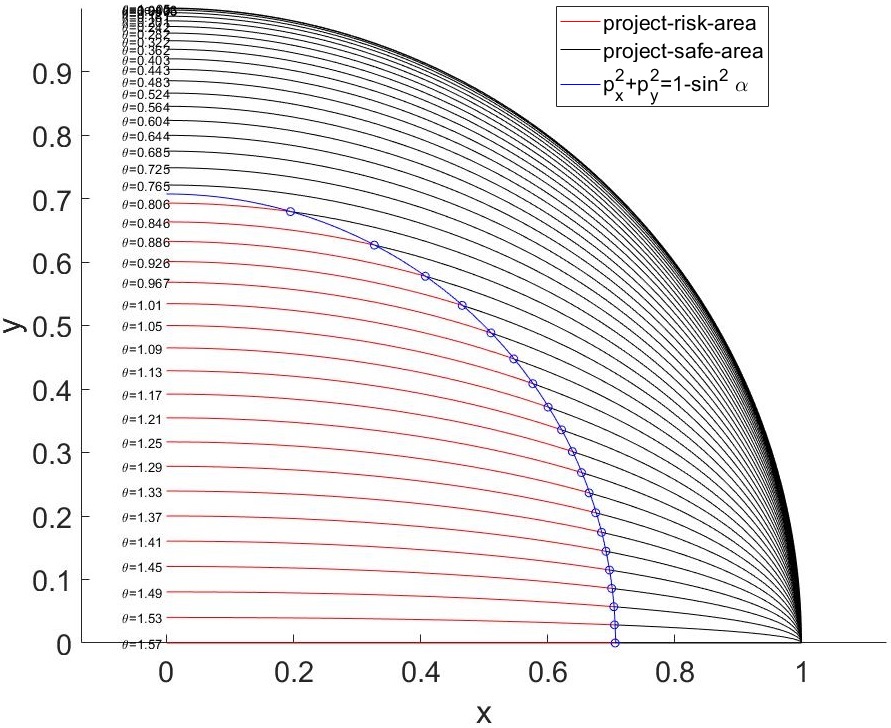}	
\caption{When $\alpha=\pi/4$, this figure shows the projected ellipse of the bottom-circle of a cylinder with different $\theta$, where the red arcs indicates the risky regions. The ratio of risky region is evaluated as the arc length ratio of the red region vs. the total elliptic arc.}
\label{fig:Appendix2}
\end{figure}

\begin{figure}
\centering
\includegraphics[width=\linewidth]{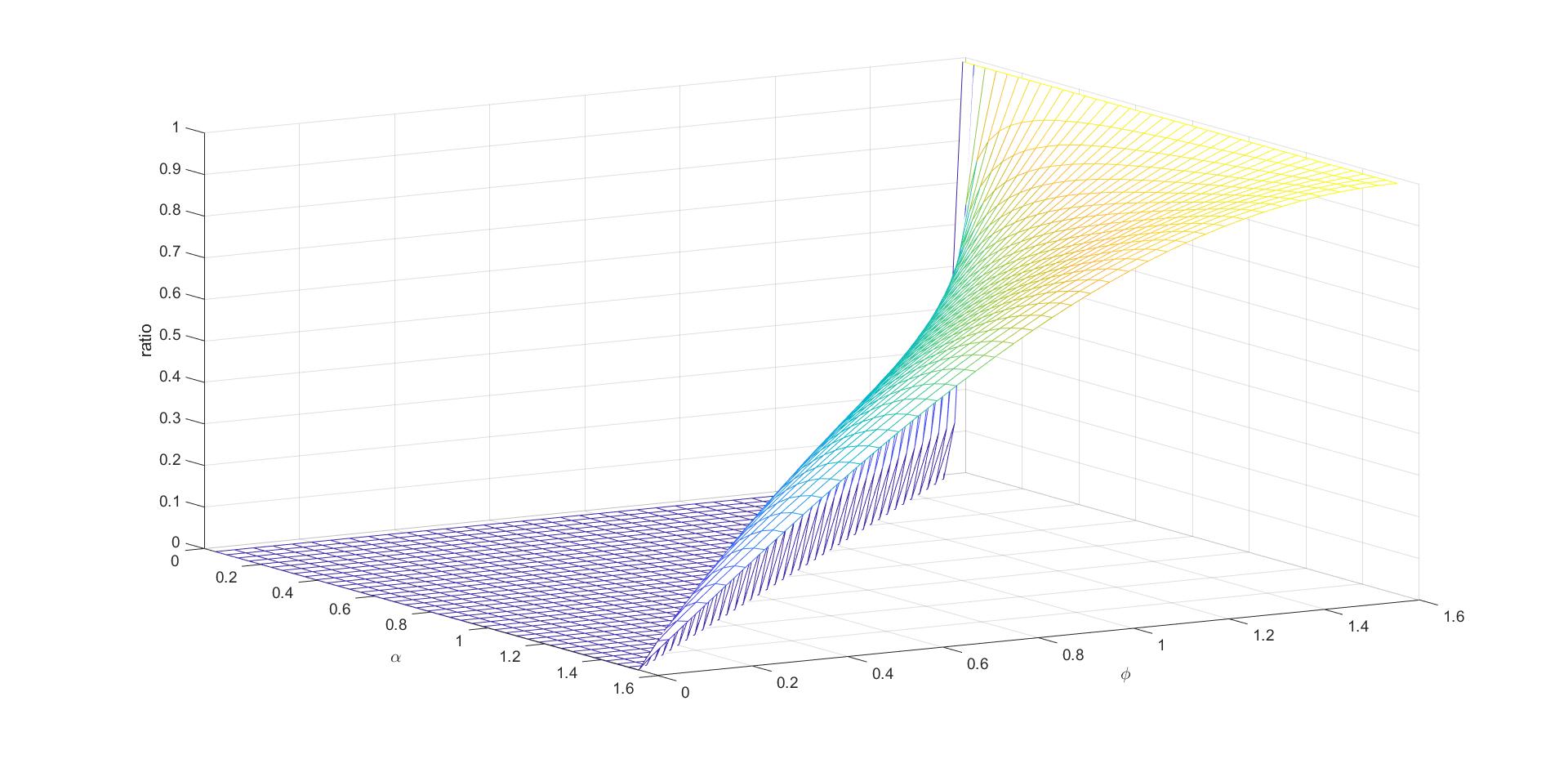}	
\caption{The 3D shape of $g(\theta, \alpha)$ as a height field of $(\theta, \alpha)$.}
\label{fig:Appendix3}
\end{figure}

The ratio of risky region for the whole projected area can be evaluated as the ratio of arc length in the region $\phi > \phi_0$. As illustrated in Fig.\ref{fig:Appendix2}, the ratio of the red curve's length on the whole ellipse is the ratio of risky region. As a consequence, we derive the following general formula for $g(\theta, \alpha)$ as
\begin{equation}
g(\theta, \alpha)=\left\{
\begin{aligned}
& 0 & & (\sin \theta \leq \sin \alpha), \\
& \frac{\int_{0}^{\phi_{0}} (\sin ^{2} \phi+\sin ^{2} \theta \cos ^{2} \phi)^{\frac{1}{2}} d \phi}{\int_{0}^{\frac{\pi}{2}} (\sin ^{2} \phi+\sin ^{2} \theta \cos ^{2} \phi)^{\frac{1}{2}} d \phi} & & (\sin \theta > \sin \alpha).
\end{aligned}
\right.
\end{equation}
The corresponding shape of $g(\theta, \alpha)$ is given in Fig.\ref{fig:Appendix3}.

To ease the computation of $g(\cdots)$ in optimization, we approximate it by polynomials as follows.
\begin{equation}
g(\theta, \alpha) \approx \left\{
\begin{aligned}
&0, & & (\sin \theta \leq \sin \alpha) \\
&\Sigma_{i=0}^5\Sigma_{j=0}^5 a_{i,j} \theta^i \alpha^j & & (\sin \theta > \sin \alpha)
\end{aligned}
\right.
\end{equation}
with
\begin{center}
$
   a_{i,j} =  10^{-1}\times \begin{bmatrix} 
                -5.15 &  31.71  & -56.03 & 34.12  & -5.45 & 0.43   \\ 
                26.36 & -127.20 & 164.41  & -67.87 & 4.99   &   0  \\
                -38.00  & 134.20  & -106.20  & 25.86     &0    &    0 \\
                17.79  &  -49.65 & 1.64   & 0      & 0    &  0  \\
                0.60  & 6.60  &  0     &0       &0     &  0 \\
                -1.66 & 0     &0        &0       &0     &0
            \end{bmatrix}.
$
\end{center}

\vfill
\end{document}